\documentclass[final,5p,times,twocolumn,authoryear,linenumber]{elsarticle}
\usepackage{amssymb}
\usepackage{float}
\usepackage{lipsum}
 \usepackage{xcolor} 
\usepackage{float} % in preamble
\usepackage{orcidlink}
\usepackage[normalem]{ulem} 
\usepackage{amsmath}
\usepackage{bm}
\usepackage[normalem]{ulem}
\usepackage{multirow}
\usepackage{array}
\usepackage{url}
\usepackage{newtxtext}
\usepackage[varvw]{newtxmath}
\usepackage{adjustbox}

\journal{High Energy Astrophysics}

\begin{document}

\begin{frontmatter}

\title{Multi Messenger Study of GRB~221009A with VHE Gamma-ray and Neutrino Afterglow from a Gaussian Structured Jet}

\author[first]{Tanima Mondal\orcidlink{0000-0002-9445-1405}}

\affiliation[first]{
  organization={Department of Physics, Indian Institute of Technology Kharagpur},
  city={Kharagpur},
  state={West Bengal},
  postcode={721302},
  country={India}
}

\author[second,third,fourth]{Soebur Razzaque\orcidlink{0000-0002-0130-2460}}
\affiliation[second]{organization={Centre for Astro-Particle Physics (CAPP) and Department of Physics, University of Johannesburg},%Department and Organization
            addressline={PO Box 524}, 
            city={Auckland Park},
            postcode={2006},
            country={South Africa}}
\affiliation[third]{organization={Department of Physics, George Washington University},%Department and Organization
            addressline={}, 
            city={Washington},
            postcode={DC 20052},
            country={USA}}
\affiliation[fourth]{organization={National Institute for Theoretical and Computational Sciences (NITheCS)},%Department and Organization
            addressline={Private Bag X1}, 
            city={Matieland},
            postcode={},
            country={South Africa}}

\author[fifth,second]{Jagdish C. Joshi\orcidlink{0000-0003-3383-1591}}
\affiliation[fifth]{organization={Aryabhatta Research Institute of Observational Sciences},
            addressline={Manora Peak}, 
            city={Nainital},
            postcode={263129},
            country={India}}
            
\author[sixth]{Sonjoy Majumder\orcidlink{0000-0001-9131-4520}}
\affiliation[sixth]{organization={Department of Physics, Indian Institute of Technology Kharagpur},
            addressline={}, 
            city={Kharagpur},
            postcode={721302}, 
            state={West Bengal},
            country={India}}

\author[seventh]{Debanjan Bose\orcidlink{0000-0003-1071-5854}}
\affiliation[seventh]{organization={Department of Physics, Central University of Kashmir},%Department and Organization
            addressline={}, 
            city={Ganderbal},
            postcode={191131}, 
            state={Jammu \& Kashmir},
            country={India}}

\begin{abstract}
Recent detections of very-high-energy (VHE; $\gtrsim 100$~GeV) emission from GRB afterglows, most notably the unprecedented brightness of GRB~221009A observed by LHAASO, reveal emission components beyond the standard electron synchrotron model. The multi-TeV photons motivate synchrotron self-Compton and possibly hadronic contributions, while the non-detection of coincident neutrinos by IceCube/KM3NeT/GRAND200k constrains the microphysical parameters, jet kinetic energy and ambient medium density. We model the VHE afterglow of GRB 221009A with an external forward shock from a Gaussian structured jet in a uniform density medium. This angular structure reproduces the extreme TeV output at an off-axis angle but without demanding large energies as in a top-hat jet. We also compute the corresponding $p\gamma$ neutrino flux in the PeV-EeV energies and derive a time-integrated upper limit based on the effective area of IceCube Gen2 and GRAND200k. This provides us with insights into the contributions of individual GRBs to the neutrino events. The predicted neutrino flux for GRB 221009A with model parameters inferred from multi-wavelength spectral energy distribution lies below the sensitivities of these detectors. Even our correlation analysis, optimized to search for neutrino signals by the upcoming  GRAND200k, infers that the events from this GRB are of order $\sim 0.1$ under a highly optimistic microphysical parameter regime. We also compare the variation in neutrino flux arising from on-axis and off-axis jet viewing geometries, and conclude that it can account for approximately an order of magnitude difference in the neutrino signal. Thus, our studies conclude that a brighter burst occurring closer than GRB 221009A would be crucial for any neutrino detections by upcoming neutrino telescopes. Future GRB detections by the Cherenkov Telescope Array will provide important constraints on their geometry, radiation mechanisms, and any potential associated neutrino signals.
\end{abstract}


\end{frontmatter}

\section{Introduction}
\label{introduction}

Gamma-ray bursts (GRBs) are one of the most energetic explosions in the Universe, and their broadband afterglow emission, from radio waves to very-high-energy (VHE) $\gamma$-rays, is well described by synchrotron radiation produced by relativistic electrons accelerated in external shocks~\citep{optical_prediction, Costa_1997Natur.783C, Van1997Natur_686V, frail1997Natur_261F, Sari1998ApJ_17S, Llyod2000ApJ.722L, burgess2020NatAs.174B}. In recent years, detections of GeV–TeV emission by ground-based Cherenkov telescopes~\citep{acciari2019magic, abdalla2019very, hess2021revealing, lhaaso2023very, abe2024magic} have revealed radiation components beyond the standard synchrotron model. Proposed mechanisms include synchrotron self-Compton (SSC) scattering, where relativistic electrons upscatter their own synchrotron photons to TeV energies~\citep{chiang1999synchrotron, dermer2000beaming, sari2001synchrotron, murase2011implications, fraija2019synchrotron, wang2019synchrotron, derishev2019physical, joshi2021modelling, mondal2023probing, ren2024jet}; proton-synchrotron emission from ultra-relativistic protons~\citep{isravel2023proton}; and photo-hadronic ($p\gamma$) interactions producing electromagnetic cascades and high energy neutrinos~\citep{razzaque2010leptonic, razzaque2013long, sahu2022very}. These processes provide a plausible probe of both the particle acceleration efficiency and the microphysical conditions in GRB blast waves, while also linking GRBs to the broader context of ultra-high-energy cosmic rays (UHECRs) and neutrinos~\citep{waxman1995cosmological,vietri1995acceleration}.

GRBs are considered as potential sources of UHECR acceleration. Detecting ultra-high energy (UHE) ($\geq$PeV) neutrinos from them would serve as a direct way to assess their role in UHECR production. Within the fireball framework of GRB jets, high energy neutrinos can be produced in multiple interaction scenarios. In internal shocks, neutrinos may arise from interactions of shock-accelerated cosmic-ray protons with prompt photons \citep{waxman1997high}. Later, in the external forward shocks of the blast wave, non-thermal protons interacting with afterglow photons can generate TeV–PeV neutrinos \citep{waxman2000neutrino, razzaque2013long, razzaque2015pev}. In addition, the external reverse shock has been proposed as an efficient site for the acceleration of UHECRs, leading to the possible production of EeV neutrinos \citep{murase2007high}.

The current generation neutrino detector, IceCube, reported more than one hundred VHE astrophysical neutrinos through the High-Energy Starting Events (HESE) catalogue \citep{aartsen2020time, abbasi2021icecube, abbasi2022searches}, along with 276 events through the IceCube Event catalogue of Alert Tracks \citep{abbasi2023icecat}. However, despite extensive stacking analyzes of hundreds of GRBs and dedicated follow-ups of exceptionally bright events such as GRB~080319B, GRB~130427A, GRB~160625B, and GRB~221009A, no significant neutrino associations have been found with GRBs, either during prompt emission or in the afterglow phases \citep{abbasi2023limits}.

On 9 October 2022, at 13:16:59 UT ($T_{0}$), the Gamma-ray Burst Monitor (GBM) on board \textit{Fermi} detected GRB~221009A, an exceptionally luminous and nearby ($z \sim 0.151$) long-duration GRB~\citep{deugarte2022_1D,veres2022grb}. The burst was subsequently observed by several instruments, including \textit{Swift}~\citep{williams2023grb}, \textit{Fermi-LAT}~\citep{lesage2023fermi,banerjee2025observation}, INTEGRAL/SPI-ACS, and Konus-Wind~\citep{frederiks2023properties}, with its localization refined through InterPlanetary Network triangulation~\citep{2022GCN.32641....1S}. Among the extensive multiwavelength and multi messenger follow-ups, a particularly remarkable discovery was the detection of photons with energies exceeding 10~TeV, reported by LHAASO \citep{lhaaso2023very, huang2022lhaaso}, marking the first ultra-high-energy $\gamma$-ray observation from a long GRB.

The large isotropic kinetic energy and relatively close distance make GRB~221009A a promising target for high-energy neutrino detection.
Being the brightest-of-all-time (BOAT) event, it offers a rare opportunity to probe neutrino production in GRB jets and to test the efficiency of hadronic processes in such extreme explosions. Several recent studies have explored the implications of IceCube’s non-detection of neutrinos from this burst by analysing its time-integrated $\gamma$-ray flux \citep{murase2022neutrinos, ai2023model}. This null result provides important constraints on the physical conditions of GRB outflows and places stringent limits on the allowed parameter space of emission models \citep{ai2023model, veres2024non}.

Different energy ranges of neutrino events require numerous analysis strategies in current detectors such as IceCube \citep{abbasi2023limits} and KM3NeT \citep{aiello2019sensitivity}, which search for neutrino emission from astrophysical transients like GRBs over broad energy and temporal intervals. For GRB~221009A, the IceCube Collaboration conducted a dedicated search using its real-time Fast Response Analysis (FRA), optimized for TeV–PeV neutrinos \citep{2022GCN.32665....1I}. The search was performed at a reference energy $E_{0}=100\,\mathrm{TeV}$ during two time windows, $[T_{0}-1\,\mathrm{hr},\,T_{0}+2\,\mathrm{hr}]$ and $[T_{0}\pm 1\,\mathrm{day}]$, and reported only upper limits on neutrino emission \citep{abbasi2023limits}. Within the three-hour window, IceCube set a 90\% confidence-level upper limit on the time-integrated muon-neutrino flux of $E^{2}\,\mathrm{d}N/\mathrm{d}E \le 3.9\times 10^{-2}\ \mathrm{GeV\ cm^{-2}}$ assuming an $E^{-2}$ power-law spectrum, with a sensitivity range of approximately $800\,\mathrm{GeV}$–$1\,\mathrm{PeV}$\footnote{\url{https://gcn.gsfc.nasa.gov/other/221009A.gcn3}}. Similarly, the KM3NeT Observatory carried out a search in the interval $[T_{0}-50\,\mathrm{s},\,T_{0}+5000\,\mathrm{s}]$ and also reported no evidence for neutrino emission \citep{ 2022GCN.32741....1K, aiello2024search}.

Several numerical studies have attempted to explain the origin of the TeV photons observed from GRB~221009A, invoking top-hat jet models with varying opening angles~\citep{sato2023two} or structured jets characterized by shallow power-law or Gaussian or two-component jet angular profiles~\citep{o2023structured, gill2023grb, zhang2024boat, ren2024jet}. Such jet structures are expected to arise naturally from jet break through the progenitor star’s envelope or from cocoon formation during collapse~\citep{lamb2017electromagnetic, gottlieb2018cocoon, lazzati2018late}, and they strongly shape the observed afterglow emission through their angular energy and velocity distributions at different viewing angles. In our recent work~\citep{mondal2025follow}, we demonstrated that a Gaussian structured jet can successfully reproduce VHE off-axis afterglow emission and explored how afterglow parameters govern SSC light curves and TeV detectability with CTA. Building on this framework, in this study, we aim to investigate whether the Gaussian structured jet scenario can also account for the TeV photons of GRB~221009A, while extending the analysis to calculate the associated neutrino flux in a constant-density interstellar medium (ISM) with an adiabatically evolving forward shock. In particular, the impact of off-axis structured jets on the resulting neutrino flux remains poorly explored across the existing literature. In this work, we systematically investigate how both intrinsic and extrinsic jet parameters influence the neutrino flux and assess its detectability with current and upcoming observatories, such as IceCube Gen2 and GRAND200k. This approach enables us to identify the correlation among parameter regimes that strongly enhance the neutrino flux and increase the number of neutrino events towards the current upper limit of those neutrino detectors. This correlation study not only constrains the baryon loading and particle acceleration efficiency in GRB jets but also provides tighter constraints on parameters associated with the neutrino flux. 

This paper is organized as follows. Section~\ref{section2} describes the blast wave afterglow model with a Gaussian structured jet in a uniform ISM and computes the VHE $\gamma$-ray emission, including synchrotron and SSC components with Klein–Nishina, internal $\gamma\gamma$ attenuation, and EBL corrections. Section \ref{section3} investigates the parameter dependence of the VHE light curves, details the fitting methodology, and applies the model to GRB 221009A through spectral modeling and TeV afterglow light-curve analysis. Section \ref{section4} formulates the afterglow neutrino emission from the $p\gamma$ channel, and the impact of jet angular structure on the flavor-mixed neutrino flux at Earth. Section \ref{section5} presents the resulting constraints, comparing the predicted neutrino fluxes with IceCube Gen2 and GRAND200k upper limits. The section further explores a correlation study on key parameters that govern neutrino production and quantifies expected event yields within the GRAND200k sensitivity band. Finally, we conclude with a summary of the main results and their implications for GRB jet structure and multi messenger detectability.

\section{VHE Afterglow from Gaussian Structured Jets}\label{section2}

In this section, we present a Gaussian structured jet model to interpret the GeV--TeV afterglow emission of the BOAT GRB~221009A, as observed by LHAASO \citep{lhaaso2023very} and AGILE GRID \citep{foffano2024theoretical}. By fitting these data, we constrain the jet structure, radiation mechanisms, and particle acceleration processes, thereby inferring the model parameters governing the VHE afterglow. Accurate modeling of the synchrotron and SSC components requires the evolution of both the nonthermal electron population and the downstream magnetic field. Previous studies \citep{mooley2018mildly, margutti2017electromagnetic} have shown that simple narrow top-hat jet models is insufficient to explain the VHE afterglow data. Their strong lateral expansion produces excessively steep post-peak declines and flux levels inconsistent with the observed late-time emission, restricting them to fitting only the early afterglow. Moreover, the top-hat assumption of uniform energy and Lorentz factor within a sharp half-opening angle, followed by an abrupt cutoff, requires unrealistically high isotropic-equivalent energies to explain the observed brightness, often exceeding plausible GRB energy budgets.
Recent study by \citet{o2023structured} also explained the multiwavelength observation of this BOAT GRB with a shallow power-law angular profile ($\epsilon(\theta)\propto \theta^{-a}$ with $a<2$), with both jet core and wing configuration, where they have shown that shallow power-law jets have the ability to avoid the energy crisis that would be caused by uniform jets (top-hat), and thus can capture the observed afterglow behaviour with a sharp jet-break. However, through their model, they have primarily explained the radio, optical-
infrared (OIR), and late-time X-ray emission of different VHE bright GRBs, along with GRB 221009A. Further \citet{gill2023grb}, also explain Xray and optical afterglow features of GRB 221009A with their shallow angular jet structure using the forward shock model. In one of our recent works, we already explored how a Gaussian Structured jet is efficient in producing VHE off-axis afterglow emission \citep{mondal2025follow}. Thus, in this work, we adopt our VHE Gaussian structured jet model within the forward-shock adiabatic framework, which provides a typical explanation for both the extreme luminosity of GRB~221009A and its spectral and temporal evolution.

\subsection{Blast wave modeling with Gaussian Structured jet}

To model the VHE afterglow of GRB~221009A, we consider an external forward-shock scenario in which the relativistic ejecta interact with a uniform interstellar medium (ISM). For the jet dynamics, we employ the Gaussian structured-jet framework of \citet{mondal2025follow}. This framework naturally accounts for the temporal evolution of the afterglow through SSC emission, the dominant radiation mechanism responsible for the observed sub-TeV photons.  

In the Gaussian structured jet model, jet energy per unit solid angle follows a Gaussian profile as \citep{lamb2017electromagnetic}, 

\begin{equation}\label{eqn1}
    \varepsilon(\theta) = \frac{d E_{k}}{d \Omega} = \varepsilon_{c} \exp\!\left(-\frac{\theta^2}{2\theta_c^2}\right),
\end{equation}  

\noindent
where $\varepsilon_c$ is the jet kinetic energy per unit solid angle at the jet core, characterized by the total kinetic energy $E_{k}$ of the jet. The jet core angle $\theta_{c}$ shapes the jet structure, and $\theta_{j,max}$ is the maximum jet half-opening angle of the sharp-edge jet surface beyond which energy drops rapidly (see Section-\ref{effect_strc_param} for details). This angle does not have a significant impact on the total flux, since emission from low latitudes remains insignificant whenever the light curve lies above the detection threshold \citep{resmi2018low}. Along with the jet angular energy profile, jet initial velocity profile $\Gamma_{0}(\theta) \beta_{0}(\theta)$ also follows a Gaussian profile \citep{resmi2018low, mondal2025follow} as ---

\begin{equation}\label{eqn2}
    \Gamma_{0}(\theta) \beta_{0} (\theta)= \eta_{c} \exp\left ( -\frac{\theta^{2}}{2\theta_{c}^{2}} \right ),
\end{equation}

Here $\Gamma_{c} \beta_{c}=\eta_{c}$ is constant, where $\Gamma_c$ is jet core bulk Lorentz factor and $\beta_{c}$ is the normalized velocity of the jet along $\Gamma_c$.

\begin{figure}[ht]
    \centering
    \includegraphics[width=0.8\linewidth]{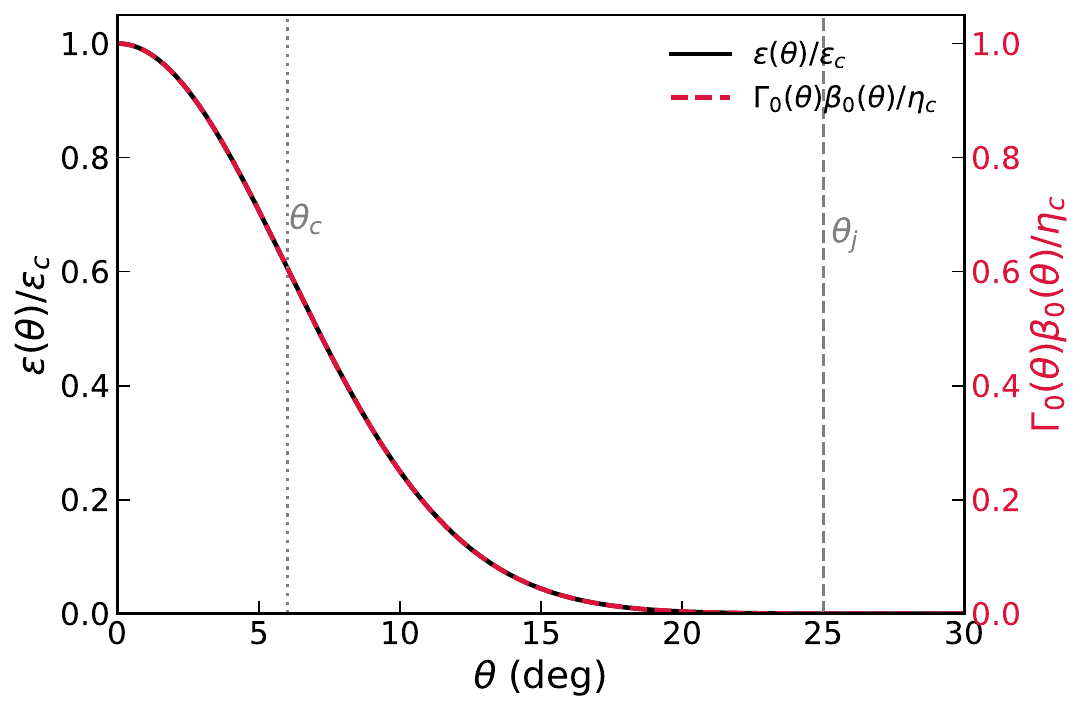}
    \caption{Normalized angular profiles of jet energy per unit solid angle, 
$\varepsilon(\theta)/\varepsilon_c$ (black solid line), and initial velocity, 
$\Gamma_0(\theta)\beta_0(\theta)/\eta_c$ (red dashed line), for a Gaussian structured jet. For example,
the jet profile is shown as a function of the polar angle $\theta$ for a jet core angle 
$\theta_c = 6^{\circ}$, jet half-opening angle $\theta_{j,\:max} = 25^{\circ}$, initial velocity $\eta_{c}=300$ and total kinetic 
energy $E_k = 1\times10^{53}~\mathrm{erg}$. Vertical dotted and dashed lines mark $\theta_c$ and $\theta_{j,\:max}$, respectively.}
    \label{fig:normalised_Eth_Gth}
\end{figure}

Figure \ref{fig:normalised_Eth_Gth} illustrates both the angular profile of the energy and the bulk Lorentz factor gradually decrease from the jet core outward $(\theta > \theta_{c})$. Furthermore, since we assume adiabatic blast wave evolution of the bulk Lorentz factor for ultra-relativistic
ejecta $(\Gamma  \beta >>1)$, the Blandford-McKee self-similar solution \citep{blandford1976fluid} provides the radial evolution of the total bulk Lorentz as $\Gamma(\theta, r)\,\beta(\theta, r) \propto 
\left( \frac{r}{r_{\mathrm{dec}}} \right)^{-3/2}$, for $r > r_{\mathrm{dec}}$. Here, $r$ is the radius of the blast wave from the center of explosion, $r_{dec}$ is the deceleration radius that can be obtained as $r_{\mathrm{dec}} = \left(3 \varepsilon_c / \eta_c^2 n_0 m_p c^2 \right)^{1/3}$ \citep{mondal2025follow}, with $n_0$ is number density of ambient medium, $m_p$ is the mass of proton, $c$ is velocity of light.

The temporal evolution of the afterglow is determined by jet intrinsic parameters --- the total kinetic energy $E_{k}$, ambient density $n_0$, jet core angle $\theta_{c}$, the energy fractions in electrons ($\epsilon_e$), magnetic fields ($\epsilon_B$) and electron power law index $k$; whereas the jet extrinsic parameters includes the viewing angle $\theta_{v}$ and the luminosity distance $d_{L}$. In contrast to the typical top-hat jet scenario, a sharp jet break is expected in the temporal evolution, once the relativistic beaming angle becomes comparable to the jet opening angle \citep{sari1999jets}. Our Gaussian structured jet model does not invoke an explicit jet-break time. The smooth angular energy distribution of the Gaussian profile naturally produces a gradual steepening of the afterglow light curve, consistent with the observed long-lived shallow decay of GRB~221009A, and alleviates the extreme energy requirements implied by a uniform jet with a late-time break \citep{o2023structured}.

\subsection{Estimation of Synchrotron and SSC flux}\label{section_2.2}

In the Gaussian structured jet scenario, the early afterglow emission is dominated 
by synchrotron radiation from shock-accelerated non-thermal electrons \citep{wijers1999physical, piran1998spectra}, which are accelerated following a characteristic power-law decay $(\propto \gamma^{-p})$ of energy distribution. In later times, sub-TeV photons are primarily produced by Synchrotron Self-Compton (SSC) scattering, where the same group of synchrotron electrons upscatter synchrotron photons to the VHE regime \citep{sari2001synchrotron}. The synchrotron spectrum is described by 
piecewise power-law segments separated by the minimum injection frequency $\nu_{m}$, the 
cooling frequency $\nu_{c}^{s}$ and maximum frequency $\nu_{M}$, associated with the electron Lorentz factors $\gamma_{m}$, $\gamma_{c}^{s}$ and $\gamma_{M}$, respectively.  Depending on the relative values of these Lorentz factors, the spectrum exhibits two cooling regimes: fast cooling ($\gamma_{m} > \gamma_{c}^{s}$) and slow cooling ($\gamma_{c}^{s} > \gamma_{m}$). At an early time of the afterglow, most of the electrons get efficiently cooled down within their dynamical timescale due to the strong magnetic field and high electron density. Whereas at a later time, when the blast wave expands, its comoving magnetic field weakens over time, and the electron cooling time dominates over the dynamical time. Hence, only the highest-energy electrons radiatively 
cool, having $\gamma > \gamma_{c}$. Furthermore, the maximum synchrotron Lorentz factor $\gamma_{M}$ is calculated by comparing synchrotron acceleration and cooling time scale, beyond which electrons can not accelerate efficiently.

The relevant Lorentz factors and associated magnetic field strength are given by

\begin{align*}
    \gamma_{m} &= \frac{m_p}{m_e}\!\left(\frac{p-2}{p-1}\right)\epsilon_{e}(\Gamma-1) + 1, \\
    \gamma_{c} &= \frac{6\pi m_{e} c (1+z)}{\sigma_{T}(1+Y) B^{2}\Gamma t_{\rm ob}}, \\
    \gamma_{M} &= \sqrt{\frac{6 \pi q_{e} e_{acc}}{\sigma_{T} B (1+Y)}},\\
    B &= \left(32\pi m_{p} c\, \Gamma^{2} \epsilon_{B} n_{0}\right)^{1/2},
\end{align*}

\noindent
where $\sigma_{T}$ is the Thomson cross-section and $B$ is the comoving magnetic field strength. 
The Compton parameter $Y$ describes the relative contribution of SSC cooling 
\citep{jacovich2021modelling, mondal2023probing}, reducing the synchrotron cooling Lorentz factor and 
modifying the cooling break frequency.

Now, to compute the synchrotron flux, for an observer at a viewing angle $\theta_v$ from the jet axis, the apparent inclination angle $\alpha_{\mathrm{inc}}^{i,k}$ of each segment relative to the line of sight is calculated following \citet{resmi2018low}, where each segments defined by polar angles $\theta_i$ and azimuthal angles $\phi_k$. Then, for those different jet segments, we calculate the deceleration radius $r_{dec}$ of the fireball. For each of the $(i,k)$th segments, observer time $t_{ob}$ is associated with its corresponding shock radius $r$ and inclination angle $\alpha_{\mathrm{inc}}^{i,k}$ \citet{resmi2018low, mondal2025follow}, i.e. $t_{ob}(r, \alpha_{inc}^{i,k}) = \frac{r}{\beta(r)c}\left [ 1-\beta(r) \cos \alpha_{inc}^{i,k} \right ]$. Thus, for a given range of r values, we have interpolated the $t_{ob}(r, \alpha_{inc}^{i,k}) - r$ equation to obtain off-axis flux contribution from each jet segment at a given $t_{ob}$. Hence, the total observed flux is obtained by summing the contributions from each jet segment at a given observer time. The observed synchrotron flux from each jet segment (i,k) is given by,

\begin{equation} 
    \begin{split}
   F^{(i,k)}_{\rm syn}(t_{ob},\alpha_{inc}^{i,k}) &= a_{dop}^{3} F_{\nu/a_{dop}}^{i,k}(a_{dop}t,\alpha_{inc}=0) \cos \alpha_{inc}^{i,k}\\ & = a_{\text{dop}}^{3}(r,\alpha_{\text{inc}}^{i,k}) \times \cos \alpha_{\text{inc}}^{i,k} \times P_{\nu, \rm max}\\
   & \times \left (\frac{n_{0}}{3}  \right ) \times r^{3} \times \frac{O_{\rm fact}}{d_{L}^{2}} \times  f_{\nu, \rm syn} \\ & = a_{\text{dop}}^{3}(r,\alpha_{\text{inc}}^{i,k}) \times \cos \alpha_{\text{inc}}^{i,k} \times F^{(i,k)}_{\rm max,syn} \times f_{\nu, \rm syn}
\end{split}
    \label{eq:general_flux}
\end{equation}

where $F^{(i,k)}_{\rm max,syn}$ is maximum synchrotron peak flux which is primarily determined by peak-averaged power per unit frequency $P_{\nu, \rm max} = m_{e}c^{2}\sigma_{T}B\sqrt{\Gamma^{2}+1}/3 q_{e}$ and the total number of electrons $N_{e}$ present in each jet segment\citep{mondal2025follow}. $f_{\nu, \rm syn}$ represents normalized synchrotron spectral profile for slow and fast cooling regimes following \citet{Sari_1998} and \citet{zhang2018physics}. Further $a_{\text{dop}}$ is the Doppler factor that accounts for the additional time to reach off-axis radiation to the observer compared to $t_{\rm obs}(\alpha=0)$ along the jet's central axis. $O_{\rm fact}$ represents the fraction of total jet solid angle associated with each discretized jet element \citep{resmi2018low}, ensuring that the total flux is obtained by summing the contributions from all angular jet segments.  For each jetted element, $O_{\rm fact}=\frac{\Omega_{i,k}}{\Omega_{e,i,k}}$, where $\Omega_{i,k}=\int_{\phi_{k-1}}^{\phi_{k}} d\phi\: \int_{\theta_{i-1}}^{\theta_{i}} d\theta \sin \theta$ is the solid angle subtended by a beamed element at a point on the central axis, and $\Omega_{e, i,k}= \rm max\left [\Omega_{i,k}, 2 \pi(1-\rm cos(1/\Gamma_{i,k}) \right ]$ \citep{resmi2018low}. The term $\cos \alpha_{\mathrm{inc}}^{i,k}$ represents accounts for the correction due to the emission area projection along the line of sight \citep{salmonson2003perspective, lamb2017electromagnetic}. Thus the above equation~\ref{eq:general_flux} represents the observed synchrotron flux contribution from each jet element, which can correspond to either on-axis ($\theta_{c}>\theta_{v}$) or off-axis ($\theta_{v}>\theta_{c}$) emission depending on the angular position of the segment.

Analogous to synchrotron emission, the SSC afterglow flux is governed by characteristic spectral break frequencies: the minimum injection frequency $\nu_{mm}^{\rm SSC} = 2\nu_m \gamma_m^2$, the cooling frequency $\nu_{cc}^{\rm SSC} = 2\nu_c \gamma_c^2$, and intermediate frequency $\nu_{mc}^{\rm SSC} = \sqrt{\nu_{mm}^{\rm SSC} \nu_{cc}^{\rm SSC}}$. The cooling regime of the electron population is determined by the relative ordering of these frequencies — slow cooling appears when $\nu_{mm}^{\rm SSC} < \nu_{cc}^{\rm SSC}$, while fast cooling occurs when $\nu_{mm}^{\rm SSC} > \nu_{cc}^{\rm SSC}$. The observed SSC flux from each jet segment can be calculated as ---

\begin{equation} \label{eqn_combined_ssc}
\begin{split}
   F^{(i,k)}_{\text{ssc}} &= F^{(i,k)}_{\text{max,syn}} \times (n_{0}r\sigma_{T}x_{0}) \times f_{\nu,\rm ssc} \\ & = F^{(i,k)}_{\text{max,ssc}} \times f_{\nu, \rm ssc}
\end{split}
\end{equation}

The SSC peak flux $F^{(i,k)}_\text{max,ssc}$ incorporates the synchrotron peak flux $F^{(i,k)}_{\text{max,,syn}}$, the blast wave radius $r$, the ambient medium density $n_0$, and the Thomson cross-section $\sigma_T$. This flux quantifies the efficiency of energy transfer from low-energy synchrotron photons to high-energy SSC photons. Thus, to obtain the afterglow flux evolution, both in synchrotron and SSC ($f_{\nu_{\rm syn}}^{(i,k)}(t)$ and $f_{\nu_{\rm SSC}}^{(i,k)}(t)$), it is crucial to estimate the time evolution of all break frequencies and peak specific flux. $f_{\nu, \rm ssc}$ is the on-axis SSC spectrum profile for slow and fast cooling regimes, which further follows analytical approximation as given in \citet{sari2001synchrotron} and \citet{gao2013compton}.

\subsection{Effect of Klein-Nishina correction, pair-production and EBL correction}

To consistently extend our VHE off-axis afterglow model, we incorporate the effects of correction for Klein-Nishina (KN) scattering cross-section to SSC emission, which becomes relevant when photon energies in the electron’s rest frame approach or exceed the electron rest mass energy, $m_e c^2$ \citep{nakar2009klein}. In this regime, the energy gain per photon per scattering becomes effectively constant. Following the prescriptions of \citet{nakar2009klein} and \citet{mondal2023probing}, we include the modified Y parameter in the KN regime, denoted as $Y_{\rm KN} = P_{\rm SSC}(\gamma)/P_{\rm syn}(\gamma)$, for both slow ($Y(\gamma_c)$) and fast-cooling ($Y(\gamma_m)$) electron distributions \citep{nakar2009klein, mondal2025follow}.

Moreover, we account for the attenuation of TeV photons due to pair production ($\gamma \gamma \rightarrow e^- e^+$) resulting from interactions with low-frequency background photons (optical, UV, and IR), as described by \citet{gould1967pair}. This absorption is modeled through an energy-dependent internal optical depth, $\tau_{\gamma\gamma}(E_\gamma)$, which is used to correct the intrinsic SSC flux at the source \citep{gould1967pair, joshi2021modelling}. Finally, we further include the attenuation of VHE photons during their propagation through the intergalactic medium due to interactions with the extragalactic background light (EBL), applying the correction factor $\exp[-\tau(E_\gamma, z)]$ following the formalism of \citet{dominguez2011extragalactic}.

\subsection{Effect of Gaussian structure jet on VHE Emission}\label{structured_jet}

In a Gaussian structured jet, the angular dependence of both the isotropic-equivalent energy distribution and the bulk Lorentz factor critically determines the production and detectability of VHE emission. Unlike the sharply bounded top-hat jet, the smooth angular profile of a Gaussian jet alleviates the severity of Doppler deboosting for off-axis observers. As a result, the observed flux is less suppressed at larger viewing angles, thereby improving the prospects of detecting significant VHE emission even for moderately misaligned lines of sight. The visibility of VHE photons is therefore strongly governed by the Doppler boosting factor through the observer viewing angle $\theta_{v}$. Further, in the process two distinct detection scenarios appear: (i) for $\theta_{v} < \theta_{c}$, corresponding to an on-axis geometry, where the observer receives intense emission from the highly relativistic jet core, with the flux peaking at early times due to strong Doppler boosting, (ii) for $\theta_{v} > \theta_{c}$, due to dominant Doppler de-boosting, the resulting emission attains lower peak flux and a delayed peak time relative to the on-axis case and (iii) mildly off-axis jet geometry where Gaussian structured jet viewed at a small but non-zero angle. Consequently, the spectral properties are highly sensitive to the ratio $\theta_{v}/\theta_{c}$. Our recent study \citep{mondal2025follow} shows that a large off-axis viewing angle (exceeding $25^{\circ}$) sharply reduces the probability of detection with CTA, while the VHE detectability is preferred for $\theta_{v}/\theta_{c} \leq 2.7$. In the following section, we investigate how these jet intrinsic and extrinsic parameters are correlated in shaping the observed VHE emission from GRB~221009A.

\section{Parameter Dependence of VHE Afterglow Emission}\label{section3}
\begin{figure*}[ht!]
    \centering
    \includegraphics[width=0.7\textwidth]{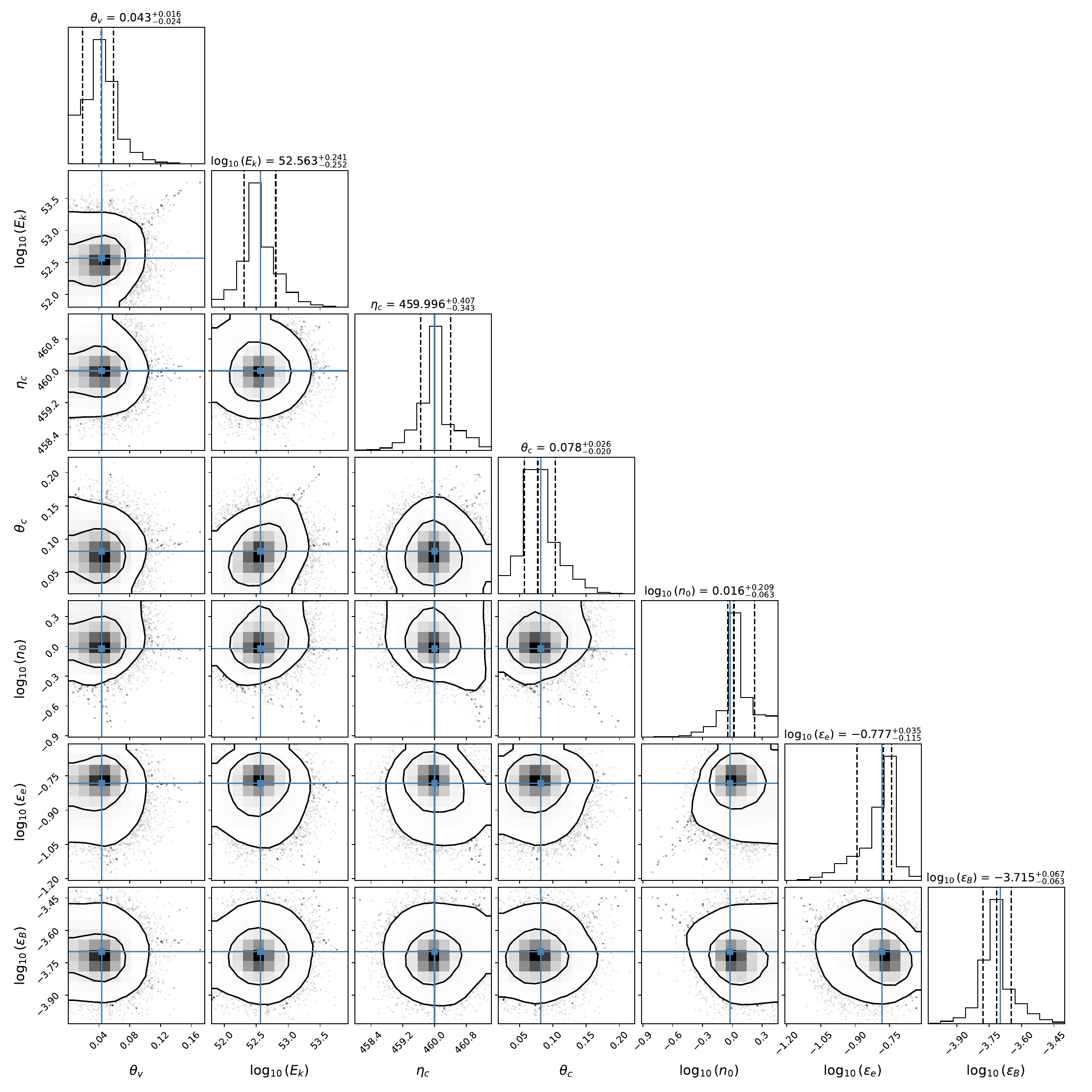}
    \caption{Corner plot showing posterior distributions of afterglow model parameters for GRB 221009A. All the contours enclosing median values and $\pm1 \sigma$ uncertainties. Thin reference lines denote the truth values.}
    \label{fig:corner}
\end{figure*}

Using the described structure jet model, we will study the SSC TeV afterglow features in both the spectral and temporal domains of GRB 221009A. Since both the peak flux and peak time depend sensitively on the underlying afterglow parameters, we perform a comprehensive exploration of the afterglow parameter space by constraining the structured jet properties that favour VHE emission from GRB 221009A. Thus, we systematically examine the spectral and temporal evolution of the afterglow through predicting spectral energy distributions (SEDs) and TeV light curves (LCs).

\subsection{Fitting Method}\label{model_parameters}
Our structured jet afterglow model is characterized by a parameter set $\Theta_{A} = {E_k, n_0, \eta_c, \theta_{j,max}, \theta_c, \epsilon_e, \epsilon_B, k, \theta_v}$, while the source luminosity distance is at $d_L = 723.6$ Mpc, and $\eta_{c}=\Gamma_{c}\beta_{c}$ is the initial velocity of jet along its core. To fit our Gaussian structured jet model to GeV-TeV data of AGILE GRID and LHASSO, we constrained our afterglow parameter space with seven parameters: $E_k, n_0, \eta_c, \theta_c, \epsilon_e, \epsilon_B, \theta_v$, while keeping $\theta_{j,\:max}$, and $k$ fixed to $25^{\circ}$ and $2.5$ respectively. Here we restrict $\theta_{j,\:max}$ value to that value ($\leq 25^{\circ}$) to avoid errors while performing numerical methods in synchrotron and subsequent SSC flux calculation. To further reduce the number of free parameters, we fix the electron power-law index k to a representative value within the range commonly inferred for GRB afterglows, which is also adopted in the neutrino-flux calculations in later (see section \ref{neutrino_flux}.

All the data of AGILE GRID and LHASSO is taken from \citet{foffano2024theoretical} and \citet{lhaaso2023tera}. We perform posterior parameter inference using the Markov Chain Monte Carlo (MCMC) technique, employing the Python package emcee\footnote{\url{https://emcee.readthedocs.io/en/stable/}} as the sampler \citep{foreman2013emcee}. This Bayesian approach enables us to explore the multidimensional parameter space efficiently and estimate posterior probability distributions for the model parameters given the observed GeV–TeV afterglow data. We adopt broad, uniform priors for all free parameters within physically motivated ranges, relevant to GRB afterglow modeling, and do not account for any systematic errors of the instruments. The likelihood function was defined assuming Gaussian uncertainties in the AGILE–GRID and LHAASO flux measurements \citep{foffano2024theoretical, lhaaso2023tera}.

\begin{table}[ht!]
\caption{Best-fit Afterglow model parameters from simultaneous interpretation of VHE light curves and SED spectrum of GRB 221009A. Uncertainties denote 1$\sigma$ credible ranges from MCMC posterior distributions.}
\label{tab:afterglow_params}
\vspace{2mm} 
\centering
\begin{tabular}{cc}
\hline
\hline
\textbf{Parameter}  & \textbf{Range}     \\ \hline
$\log E_k(\mathrm{erg})$    & $52.66^{+0.24}_{-0.25}$          \\[1.5mm]
$\log n_{0}$ (cm$^{-3}$) & $0.02^{+0.21}_{-0.06}$
\\[1.5mm]
$\eta_{c}$ & $459.56^{+0.41}_{-0.34}$
\\[1.5mm]
$\log \epsilon_{e}$     & $-0.78^{+0.04}_{-0.12}$          \\[1.5mm]
$\log \epsilon_{B}$     & $-3.71^{+0.07}_{-0.06}$          \\[1.5mm]
$\theta_{v}$ (deg)  & $2.46^{+0.92}_{-1.37}\,^{\circ}$ \\[1.5mm]
$\theta_{c}$ (deg) & $4.41^{+1.49}_{-1.14}\,^{\circ}$  \\
\hline
\end{tabular}
\end{table}

 Initially, to fit the spectra of GRB 221009A, among the three time intervals of the SED (see Section~\ref{SEDs}), the MCMC sampling was performed over the range $T^{\star} + [100,674]$~s. $T^{\star} = T_0 + 226$~s is the reference time considered in most spectral analyses of GRB~221009A \citep{lhaaso2023tera, foffano2024theoretical, Ren_2024, abe2025grb}. The resulting posterior probability densities were obtained using 32 walkers in emcee, each of which was evolved for 2000 iterations to ensure convergence. The parameters of the best-fit model along with their associated confidence intervals ($\pm 1\sigma$) are summarized in Table~\ref{tab:afterglow_params}, and the corresponding posterior distributions are illustrated in the corner plot (Figure~\ref{fig:corner}). To extend the analysis to the other two SED intervals, we allow the fractional magnetic field energy $\epsilon_B$, which represents the fraction of post-shock energy transferred in magnetic fields, to evolve as a function of observer time under adiabatic expansion in the ISM, while keeping all other parameters within their $\pm 1\sigma$ credible intervals. This time-dependent treatment yields a consistent fit to the broadband SEDs and provides good agreement with the GeV–TeV light-curve data of GRB~221009A observed by LHAASO and AGILE–GRID.

\subsection{VHE Spectral modelling of GRB 221009A}\label{SEDs}

\begin{figure}[ht!]
  \centering
  \includegraphics[width=0.8\linewidth]{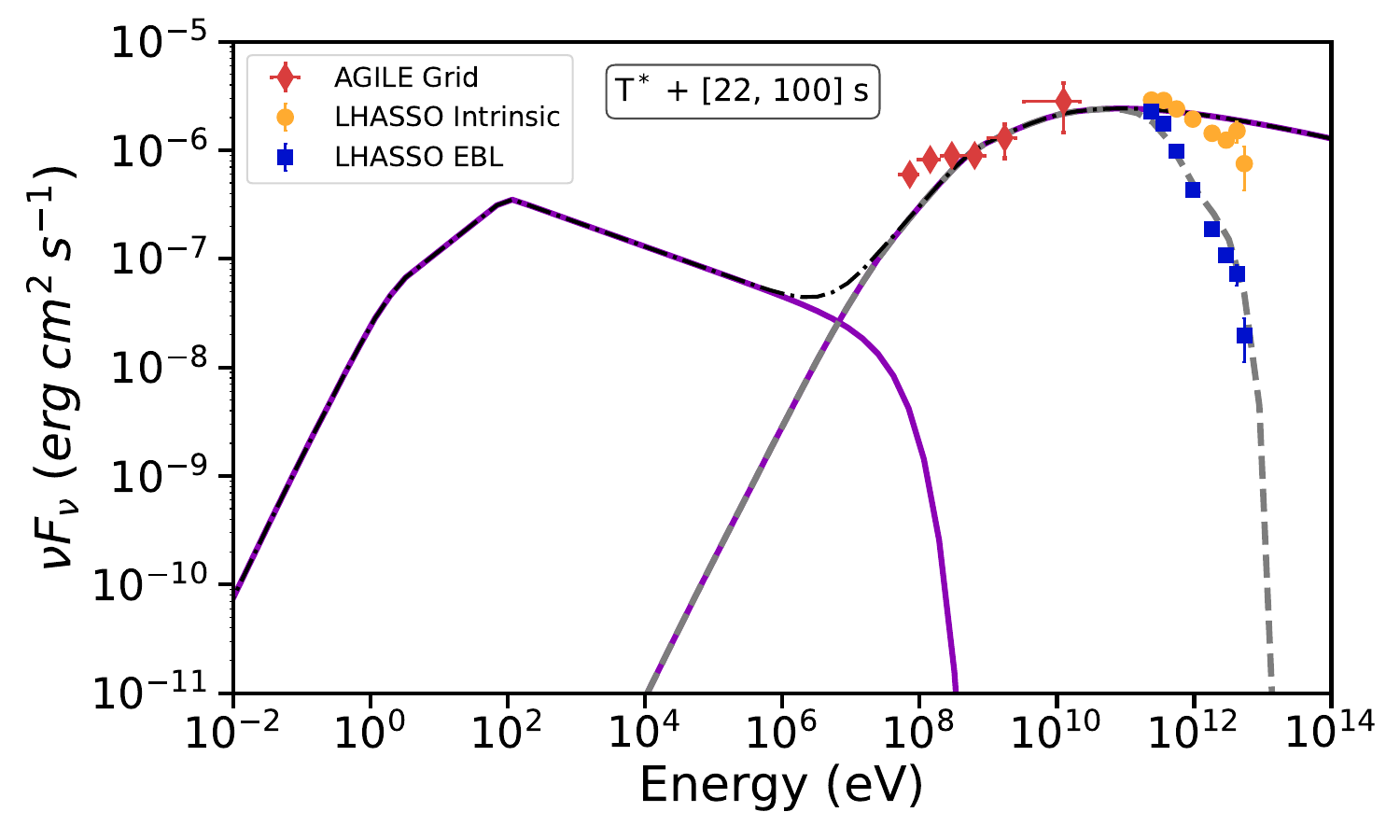}\hfill
  \includegraphics[width=0.8\linewidth]{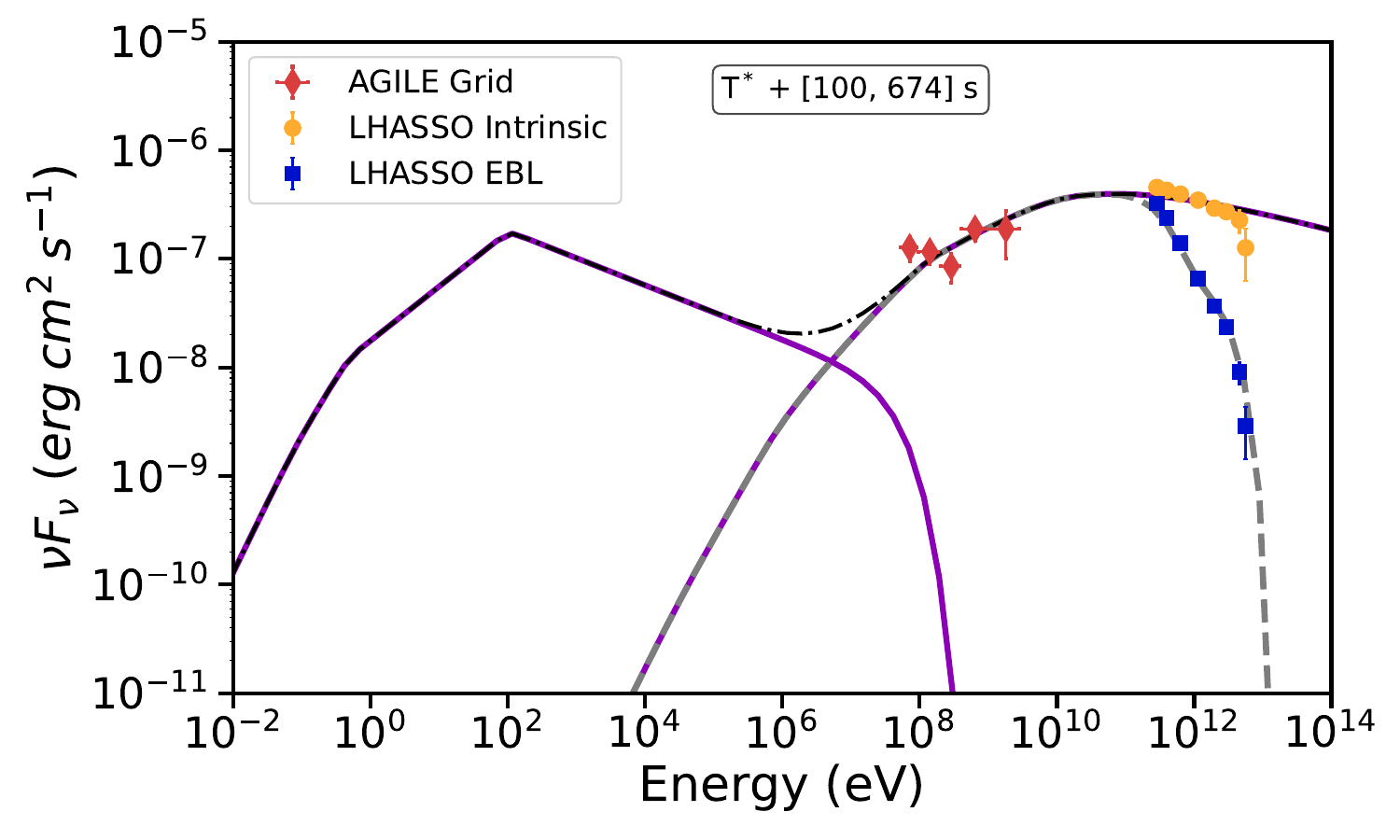}\hfill
  \includegraphics[width=0.8\linewidth]{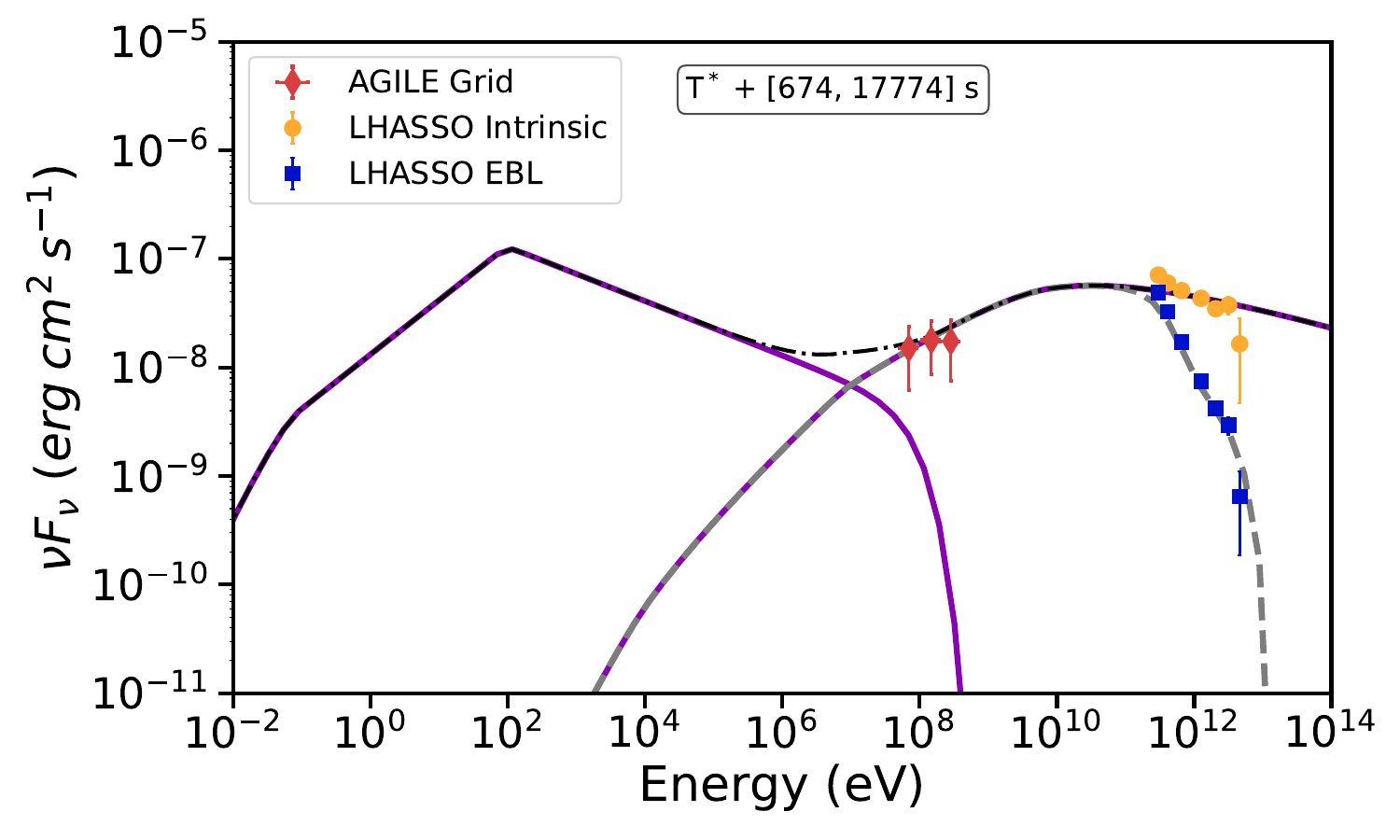}
  \caption{Broadband SEDs of GRB 221009A at three time intervals derived from the structured-jet afterglow model. Top: $T^{*}+[22,100]$ s; middle: $T^{*}+[100,674]$ s; bottom: $T^{*}+[674,1774]$ s, where $T^{*}=T_{0}+226$ s. Red diamonds denote AGILE–GRID GeV data, orange circles show LHAASO intrinsic TeV fluxes, and blue squares include EBL attenuation. The solid purple curves represent the intrinsic model spectra, with the low-energy hump from synchrotron emission and the high-energy hump from SSC radiation. The dashed grey lines show spectra corrected for EBL attenuation, while dash-dotted black lines indicate the total synchrotron + SSC emission.}
  \label{Three-SEDs}
\end{figure}

\begin{figure}[ht!]
    \centering
    \includegraphics[width=0.8\linewidth]{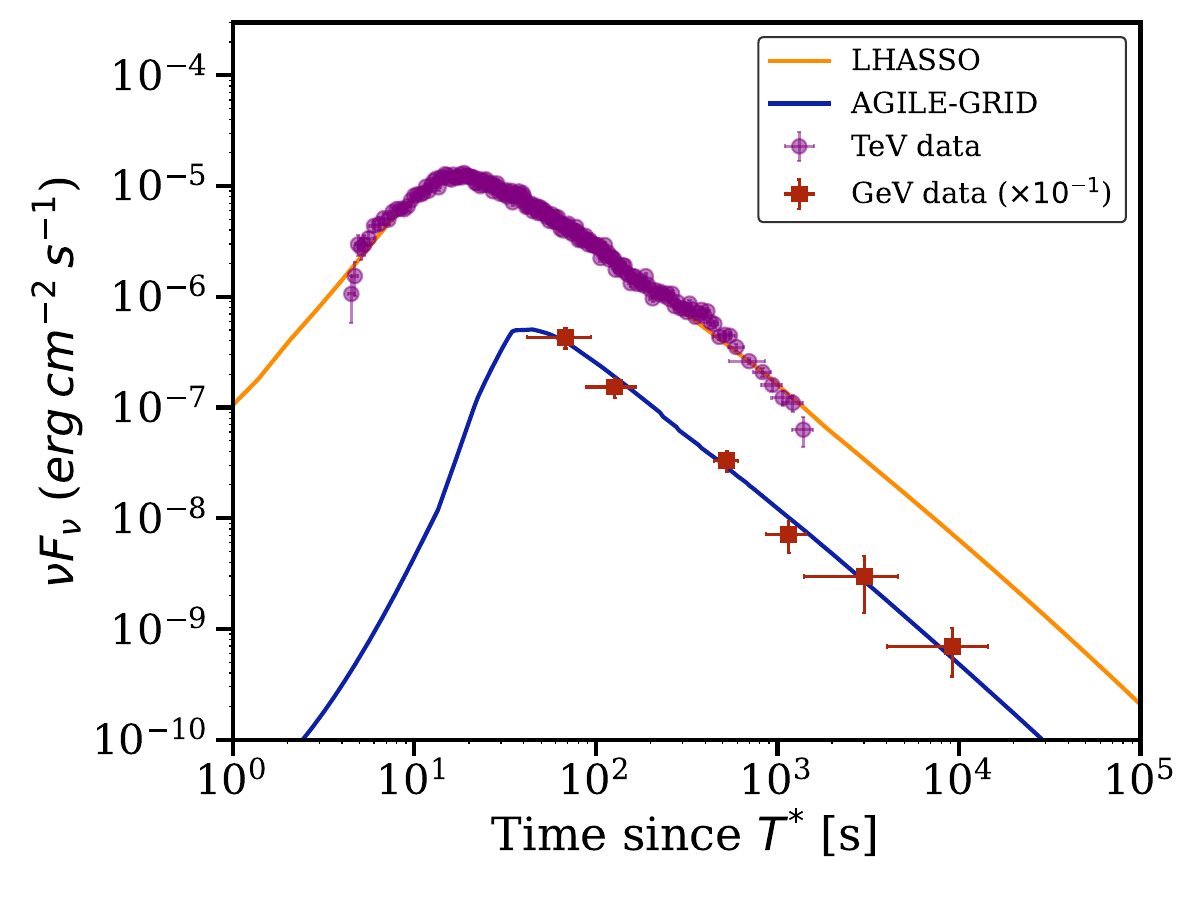}
    \caption{Afterglow light curves of GRB 221009A in the GeV and TeV bands modeled with the Gaussian structured-jet framework. The orange curve represents the TeV LC, fitted with 
    LHAASO TeV data for 0.3–5 TeV, and the blue curve shows the GeV LC fit to AGILE–GRID GeV data of 50 MeV–3 GeV. Purple circles correspond to the observed LHAASO TeV data, while dark-red squares denote AGILE–GRID GeV data (scaled by $\times 10^{-1}$).}
    \label{fig:LC_AGILE}
\end{figure}

In this section, we validate our structured jet model by globally fitting the early afterglow GeV-TeV spectral data of GRB 221009A from AGILE and LHASSO. As AGILE GRID GeV data are not available for the very early afterglow phase of the GRB, specifically $T^* + [5, 14]$ s and $T^* + [14, 22]$ s. In our analysis, we have mainly used the GeV and TeV spectral data of the time bin of $T_{1} = T^* + [22,100]$ s, $T_{2}=T^* + [100,674]$ s and $T_{3}=T^* + [674,1774]$ s \citep{foffano2024theoretical}. Figure \ref{Three-SEDs} represents the optimized synchrotron and SSC spectra of GRB 221009A for these three distinct time intervals. The spectra incorporate the Klein-Nishina effect, corrections due to $\gamma \gamma$-pair production optical depth, and the EBL attenuation of TeV photons.

SEDs of all of these time intervals collectively comprise the rising, peaking, and decaying phases of the afterglow. In these SED panels, all the first spectral humps, rising from low energy, ranging from $\sim10^{-2}$ eV up to $\lesssim1$ GeV, are dominated by synchrotron emission, while the second humps correspond to SSC radiation. Solid lines represent intrinsic model spectra, dashed lines show the EBL attenuated spectra and $\gamma\gamma$ annihilation corrected spectra, and dashed-dot lines depict the combined synchrotron and SSC components. In particular, the afterglow flux begins to be dominated by the SSC around $1$ GeV, and significant EBL attenuation becomes apparent above $\sim10$ TeV.

Our model fits the observed data at the jet observer viewing angle $\theta_{v}\; \geq 0.6^{\circ}$, and the spectra lie in the regime of $\theta_{c} > \theta_{v}$ at all intervals. These parameter values lead to a mildly off-axis jet scenario, enhancing the SSC flux due to Doppler boosting. This result is in good agreement with the structured-jet interpretation proposed by \citet{o2023structured}. Furthermore, we found that all of the spectra are well explained in $\epsilon_{e}>>\epsilon_{B}$, indicating that inverse Compton cooling dominates over magnetic energy losses, thereby amplifying the SSC component. The best-fit spectral parameters derived from this analysis are subsequently employed to calculate the corresponding neutrino flux in section~\ref{section4}. To further associate this energy with the neutrino flux calculation, we need to estimate the isotropic jet kinetic energy $E_{k, \rm iso}$ along a given direction of observer viewing angle $\theta_{v}$. The detailed calculation is discussed in Section \ref{effect_strc_param}.

\subsection{TeV afterglow Light Curve}

Figure \ref{fig:LC_AGILE} presents the afterglow light curve of GRB 221009A, which is obtained through our model by fitting GeV and TeV gamma-ray data from AGILE GRID and LHASSO, respectively.
The TeV light curve of GRB 221009A is estimated in the energy range of $0.3-5$TeV, while the GeV one is calculated in the range of $50\text{MeV} - 3\text{GeV}$. The Light curves are fitted using the afterglow parameters listed in Table~\ref {tab:afterglow_params}. Here, the VHE photons are minimally Doppler de-boosted through a mildly off-axis viewing geometry. 

The light curve exhibits a two-phase structure, characterized by a gradual rise to the peak followed by a steep decay. The observed TeV light curves include KN corrections \citep{nakar2009klein, mondal2025follow}, pair-production optical depth \citep{joshi2021modelling} and an account for attenuation due to extragalactic background light (EBL) \citep{dominguez2011extragalactic}. We find that a higher kinetic energy ($E_{k}$), ambient density($n_{0}$), and bulk Lorentz factor at the jet's core ($\Gamma_c$) strongly influence the observed SSC flux by increasing the seed photon population for inverse Compton scattering. For both GeV-TeV cases, a mildly off-axis structured jet scenario allows the observer to receive intense, early-time emission from the jet core due to its moderately high Lorentz factor. Furthermore, these GeV-TeV light curves also follow $\epsilon_e > \epsilon_B$, which is crucial for efficient SSC production in the forward shock afterglow model. Note that a feature around 600 s in the TeV light curve was previously interpreted as a jet break by LHAASO Collaboration et al. (2023). However, the GeV light curve in Figure 4 does not show such a break. Therefore, we do not assume a jet break during $T \lesssim 10^4$ s in our analysis. This is consistent with conclusions drawn by \citep{geng2025spreading, kusafuka2025tev, barnard2025modelling}.

\section{Neutrino production during Afterglow Phase}\label{section4}

In this work, we focus on the ISM case, where the adiabatic forward shock is considered one of the most promising sites for ultra-high-energy cosmic-ray (UHECR) acceleration and the subsequent production of high-energy neutrinos in the PeV–EeV range. In our analysis, we compute the neutrino spectrum arising from photo-hadronic ($p\gamma$) interactions between accelerated protons and target photons within the shocked region. Our present calculation accounts only for neutrinos produced via these $p\gamma$ interactions, while the secondary electrons and $\gamma$-rays generated in these processes have not yet been followed through their cascade development. Incorporating these secondary cascades would provide a more comprehensive picture of the multi messenger emission and represents an important direction for future work.

\subsection{Neutrino flux calculation from $p\gamma$ interaction channel}

Nonthermal protons interacting with seed photons produce pions through photo-hadronic processes, which subsequently decay into secondary particles and neutrinos. Neutral pions decay into two $\gamma$-ray photons, while charged pions decay predominantly into a muon and a muon neutrino (or antineutrino) ($p \gamma \rightarrow \pi^{\pm} \rightarrow \mu^{\pm}+\nu_{\mu}(\bar{\nu_{\mu}}) $). The resulting positively (negatively) charged muons undergo a three-body decay and yield electrons (positrons) together with neutrinos (electron and muon types): ($\mu^{\pm}\rightarrow e^{\pm}+\nu_{e}(\bar{\nu_{e}})+\nu_{\mu}(\bar{\nu_{\mu}})$). This decay chain generates an initial flavor ratio of $\nu_{e}:\nu_{\mu}:\nu_{\tau}\sim (1:2:0)$ at the source, while this ratio changes to $\nu_{e}:\nu_{\mu}:\nu_{\tau}\sim(1:1:1)$ at Earth after oscillation, which is independent of the neutrino energy.

In this work, we calculate the neutrino flux for the three neutrino flavors $\nu_{\mu}$, $\nu_{e}$, and $\nu_{\tau}$ at Earth, taking into account the effects of the neutrino oscillations. We estimate neutrino fluxes associated with both on-axis and off-axis Gaussian structured jets from a GRB event at a cosmological redshift of $z=0.151$ under an extreme energetic scenario. In addition, we derive the time-integrated upper limit sensitivity curves for this point-like source based on the planned next-generation high-energy neutrino detectors, IceCube Gen2 and GRAND200k. Further using our best-fit model parameters from the VHE $\gamma$-ray leptonic model (section~\ref{model_parameters}), we compute the neutrino flux for GRB~221009A. Later, we explore how correlations among model parameters govern the neutrino flux, and also how their extreme combination leads the number of muon neutrino events towards the  $90\%$ confidence upper limit~\citep{gehrels1986confidence} of detection for next-generation neutrino detectors.

\subsection{Neutrino flux from pion and muon decay channel} \label{neutrino_flux}

To calculate the neutrino flux arising from the photo-meson decay channel in GRBs, we consider the $p\gamma$ ineteraction formalisms of \citet{razzaque2013long}, where nonthermal protons accelerated in the external forward shock following a power law spectrum $dN/dE_{p} \propto E_{p}^{-2}$ and interacting with target synchrotron photons in an adiabatic interstellar medium. Here, $E_p$ is the proton energy, and later we used $E_{\pi}$ and $E_\nu$, which are the pion energy and energy of neutrino, respectively. The resulting neutrino flux observed at Earth depends on several factors --- the photon number density $n(E_{\gamma})$, the photo-pion production optical depth $\tau_{p\gamma}(E_{p})$, and the cosmic-ray proton flux $\phi_{p}(E_{p})$ \citep{razzaque2013long}. The charged pion flux $\phi_{\pi}(E_{\pi})$ is determined from the interaction optical depth $\tau_{p\gamma}(E_{p})$ and proton flux $\phi_{p}(E_{p})$, from which the subsequent decay chains yield muons and neutrinos.

The muon neutrino flux at the Earth
$\phi_{\nu_\mu}(E_\nu)$ from pion decay, and electron neutrino and muon anti-neutrino flux $\phi_{\nu_e}(E_\nu)$, and $\phi_{\bar\nu_\mu}(E_\nu)$ from muon decay are calculated from \citet{razzaque2013long} using standard convolution integrals involving appropriate decay scaling functions therein.
%(see ~\ref{app:nu_kernels}). 
These preliminary flux calculations do not include neutrino oscillations. However, these calculations typically adopt energy fraction mappings such as $E_\nu \approx 0.2\,E_p$ for the pion channel, reflecting the typical partitioning of energy between the parent proton and the resulting neutrinos \citep{razzaque2013long, adriani2025constraining}. Now, to estimate the neutrino flux at Earth after taking into account neutrino oscillation over the astrophysical distance, there should be three flux components: $\nu_e + {\bar\nu}_e$, $\nu_\mu + {\bar\nu}_\mu$ and $\nu_\tau + {\bar\nu}_\tau$. Using the contribution from neutrino flavor mixing parameters due to neutrino oscillations over astrophysical distance, we further calculated neutrino fluxes of all three flavors --- $\phi_{\nu_e + {\bar\nu}_e}^\oplus$, $\phi_{\nu_\mu + {\bar\nu}_\mu}^\oplus$, and $\phi_{\nu_\tau + {\bar\nu}_\tau}^\oplus$. For any $\nu+{\bar\nu}$ flavor, $\phi_{\nu + {\bar\nu}}^\oplus$ is the neutrino flux at Earth after oscillations.
%\textcolor{red}{Any citation required?}

All these neutrino fluxes of all three flavors are primarily detected by a set of parameters that strongly govern the $p\gamma$ hadronic channel for neutrino emission. These parameters are--- isotropic-equivalent kinetic energy ($E_{k,\rm iso}$), number density of the circumburst ($n_0$), initial bulk Lorentz factor $(\Gamma_{0} \equiv \eta_c)$, the fraction energy transfered to electron ($\epsilon_e$), the magnetic field fraction ($\epsilon_B$), fraction of kinetic energy imparted to non-thermal protons ($\epsilon_p$), and electron power-law index $(k)$. Based on this
parameter space, we will now explore how the neutrino flux
varies and how their correlation leads to estimating the neutrino
events at the $90\%$ C.L.

\subsection{Effect of structured-jet parameters on neutrino flux}\label{effect_strc_param}

One of the key structured-jet parameters that governs the estimation of high-energy neutrino flux is the isotropic-equivalent kinetic energy, $E_{k,\mathrm{iso}}$, of the blast wave. In our off-axis Gaussian structured-jet model, the angular distribution of jet energy is described by $\varepsilon(\theta)$, which is characterized by the jet kinetic energy per unit solid angle $\varepsilon_{c}$. This $\varepsilon_{c}$ is obtained from the jet total kinetic energy $E_{k}$, which is calculated by integrating the distribution of angular energy profile over the total solid angle $d \Omega$ within the jet surface in between $0 \leq \theta \leq \theta_{j,\:max}$ as following,  

\begin{equation}
E_k = 2\pi \int_{0}^{\theta_{j,\:max}} \varepsilon(\theta)\, \sin\theta \, d\theta .
\label{eq:Ek_int}
\end{equation}

For small jet opening angles ($\sin\theta \simeq \theta$), this simplifies to  
\begin{equation}
E_k = 2\pi \varepsilon_c \int_{0}^{\theta_{j,\:max}} \theta \, 
\exp\!\left(-\frac{\theta^2}{2\theta_c^2}\right) d\theta ,
\label{eq:Ek_small_angle}
\end{equation}
The integration yields the normalized energy distribution, and jet kinetic energy per unit solid angle at the core $\varepsilon_c$ can be expressed in terms of the total jet energy $E_k$ as  
\begin{equation}
\varepsilon_c \simeq 
\frac{E_k}{\pi \theta_c^2 \left( 1 - e^{-\theta_{j,\:max}^2 / \theta_c^2} \right)} .
\label{eq:eps_c}
\end{equation}

The isotropic-equivalent kinetic energy $E_{k,iso}$ perceived by an observer located at a viewing angle $\theta_v$ depends on the angular energy profile of the jet. Instead of assuming a single line-of-sight value, we compute the effective kinetic energy within a narrow cone of angular width $\Delta\theta$ centred on $\theta_v$, such that 

\begin{equation}
E_k(\theta_v) = 2\pi \varepsilon_c 
\int_{\theta_v - \Delta\theta}^{\theta_v + \Delta\theta} 
\theta \, \exp\!\left(-\frac{\theta^2}{2\theta_c^2}\right) d\theta ,
\label{eq:Ek_v}
\end{equation}
where $\Delta\theta \approx 1/\Gamma(\theta_v,t)$ specifies the angular width of the region effectively visible to the observer at time $t>t_{\rm dec}$ due to relativistic beaming. This formulation presents a more feasible scenario in which the structured jet emission contributes to the observed flux and, consequently, to neutrino production.

The isotropic-equivalent kinetic energy along the observer’s line of sight is then given by  
\begin{equation}
E_{k,\mathrm{iso}}(\theta_v) = 4\pi\, \bar{\varepsilon}(\theta_v),
\label{eq:Ek_iso}
\end{equation}
where the effective energy per unit solid angle is
\begin{equation}
\bar{\varepsilon}(\theta_v) = \varepsilon_c 
\exp\!\left(-\frac{\theta_v^2}{2\theta_c^2}\right).
\label{eq:eps_theta_v}
\end{equation}

This approach effectively connects the angular energy structure of the jet with the observer’s geometry, enabling the estimation of $E_{k,\mathrm{iso}}$ that directly influences the predicted high-energy neutrino flux. Due to the dependency on jet geometry, the angular profile of jet velocity also becomes $\Gamma_{0}(\theta_{v})\beta_{0}(\theta_{v})$. The derived $E_{k,\mathrm{iso}}(\theta_v)$ is subsequently used in our photo-hadronic ($p\gamma$) interaction framework for further calculations, to study parameter-space correlations and to estimate neutrino event numbers for a given neutrino detector.

\subsection{Results}\label{Neutrino_all_results}

\begin{figure}[h!]
    \centering
    \includegraphics[width = 0.8\linewidth]{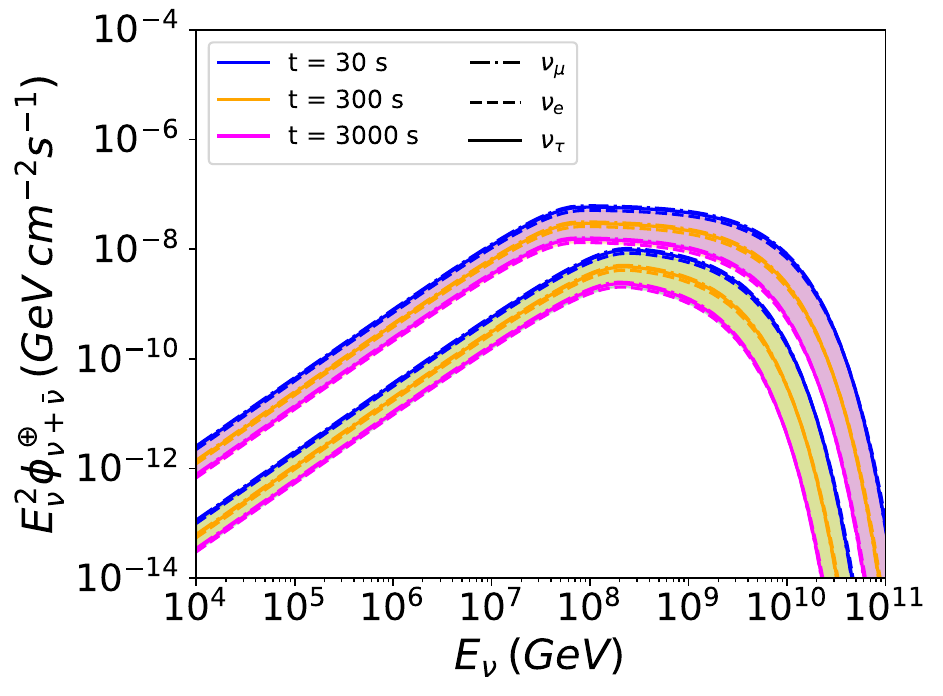}
   \caption{Neutrino flux for a GRB at $z=0.151$ in two viewing geometries. The upper (purple) band corresponds to the on-axis case ($\theta_{v}<\theta_{c}$; here $\theta_{c}=6^{\circ}$, $\theta_{v}=3^{\circ}$), while the lower (yellow) band shows the mildly off-axis case ($\theta_{v}>\theta_{c}$; $\theta_{v}=13^{\circ}$). Within each band, all three neutrino flavor fluxes at Earth after oscillation are plotted for the three observer time epochs $t > t_{dec} \in \{30,\,300,\,3000\}\,\mathrm{s}$: $\nu_{\mu}+\bar{\nu}_{\mu}$ (dash–dotted curve), $\nu_{e}+\bar{\nu}_{e}$ (dashed curve), and $\nu_{\tau}+\bar{\nu}_{\tau}$ (solid curve). Model parameters: $E_{k}=3.1\times10^{53}\,\mathrm{erg}$ (implying $E_{k,\mathrm{iso}}^{\rm on}\sim 1\times10^{56}\,\mathrm{erg}$ and $E_{k,\mathrm{iso}}^{\rm off}\sim 1\times10^{55}\,\mathrm{erg}$), $\epsilon_{e}=\epsilon_{B}=0.1$, $\epsilon_{p}=1.0$, $k=2.5$.}
   %, and initial Lorentz factors $\Gamma_{0}^{\rm on}=494$ and $\Gamma_{0}^{\rm off}=76$

    \label{Neutrino_flux_onoff_axis}
\end{figure}

\begin{figure}[h!]
    \centering
    \includegraphics[width = 0.8\linewidth]{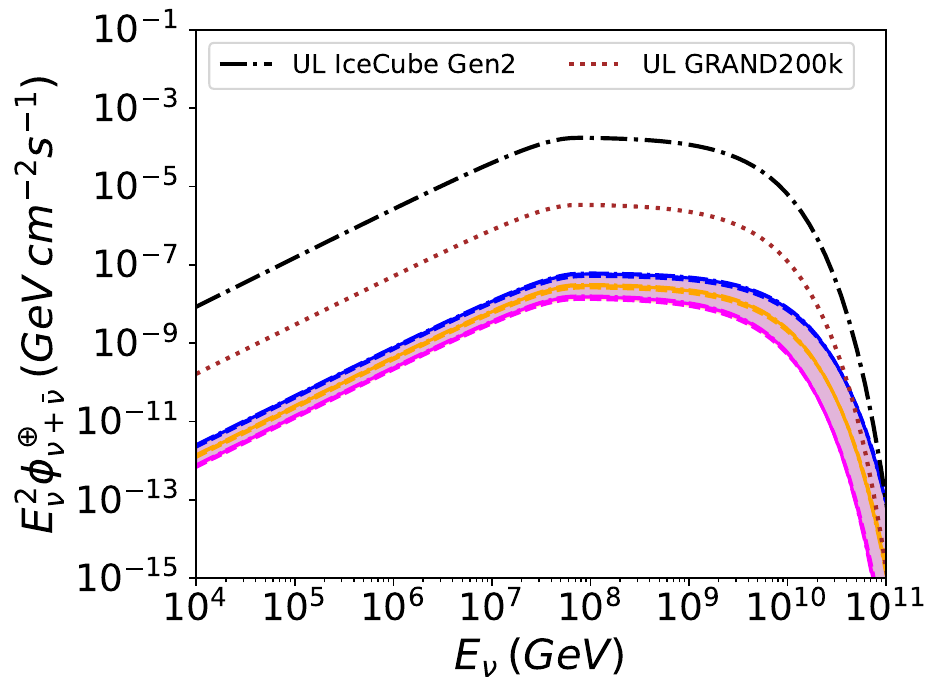}
    \caption{Time-integrated 90\% upper limit sensitivity curve of neutrino spectrum is plotted for $\nu_{\mu}$ events at $z=0.151$, for the detector geometry of IceCube Gen2 (dash–dotted curve) and GRAND200k (dotted curve). The purple band shows the on-axis neutrino flux from Figure~\ref{Neutrino_flux_onoff_axis}.}
    \label{Neutrino_flux_UL}
\end{figure}

Figure~\ref{Neutrino_flux_onoff_axis} presents the neutrino flux,
$E_{\nu}^{2}\phi_{\nu + {\bar\nu}}^\oplus$, observed at Earth after neutrino oscillation taken into account for a simulated GRB event located at redshift $z=0.151$. This figure compares two geometrical scenarios: (a) the on-axis case
$(\theta_{v}<\theta_{c})$ -- upper purple band and (b) the mildly off-axis case $(\theta_{v}>\theta_{c})$ -- lower yellow band,
with the jet core angle fixed at $\theta_{c}=6^{\circ}$, and the viewing angle
set to $\theta_{v}=3^{\circ}$ for the on-axis and $\theta_{v}=13^{\circ}$ for the mildly
off-axis scenario. Blue, orange, and magenta curves represent neutrino flux estimated at successive time epochs at $t>t_{dec}\in 30s, \:300s,\: 3000s$, respectively, illustrating the temporal decline in flux and $t_{dec}$ is the deceleration time scale~\citep{razzaque2013long}. The upper purple band corresponds to the on-axis neutrino
spectrum, displaying the summed fluxes of
$\nu_{\mu}+\bar{\nu}_{\mu}$ (dash–dotted),
$\nu_{e}+\bar{\nu}_{e}$ (dashed), and
$\nu_{\tau}+\bar{\nu}_{\tau}$ (solid), with neutrino oscillation effects included. The lower yellow band shows the corresponding off-axis
neutrino spectra for the same neutrino flavor eigenstates and time epochs. The on-axis flux is higher than the off-axis scenario, reflecting enhanced Doppler boosting and a brighter photon
target field for $\theta_{v}<\theta_{c}$, which increases the overall neutrino production.  In comparison, for off-axis geometries, the neutrino flux is suppressed due to the off-axis angular structure of the jet and the effects of Doppler de-boosting, which collectively reduce the flux observed at Earth.

As individual GRBs are effectively point sources for high-energy neutrino telescopes, it is essential to estimate the upper limit sensitivity for such sources, as this value can vary depending on the detector's geometry specification. To evaluate these limits, we utilize the effective areas of IceCube-Gen2\footnote{https://icecube-gen2.wisc.edu/science/publications/tdr/} and GRAND200k~\citep{alvarez2020giant}. The number of detected neutrino events for a single neutrino flavor over a finite time window and energy range from a given GRB by a specific neutrino detector has been calculated as ---

\begin{equation}\label{N_evnt}
    N_{\mathrm{evt}}
= \int_{T_{1}}^{T_{2}}\!\!\int_{E_1}^{E_2}
\phi_{\nu + {\bar\nu}}^\oplus\, A_{\rm eff}(E)\,\mathrm{d}E\,\mathrm{d}t \, ,
\end{equation}

\noindent
where $\phi_{\nu + {\bar\nu}}^\oplus = {\rm d}^{2}N/({\rm d}E\,{\rm d}t)$ is the differential neutrino flux at Earth and $A_{\rm eff}(E)$ is the
effective area of a specific neutrino detector for the relevant neutrino flavor.

For estimating the flux upper limit, we adopt the 90\% C.L. Poisson upper bound on the number of events, $N_{\mathrm{evt}}^{90} = 3.890$ \citep{gehrels1986confidence}. For a chosen interval $ [30, 3000]~\mathrm{s}$ and given the effective area of IceCube Gen2 and GRAND200k, we first calculate the number of muon neutrino events within the specified energy and time window using equation~\ref{N_evnt}. The normalization factor $\phi_{0}$ is then defined as the ratio of $N_{\mathrm{evt}}^{90}$ to the estimated total number of muon neutrino events for that interval. Finally, the product $\phi_{0}\,E_{\nu}^{2}\phi_{\nu}^\oplus(E)$ gives the corresponding 90\% C.L. upper limit (UL) sensitivity curve for a point-like source, consistently constraining the flux. The purple band of Figure~\ref{Neutrino_flux_UL} shows the on-axis neutrino flux from Figure~\ref{Neutrino_flux_onoff_axis}. The resulting time-integrated upper limit curve ($90\%$ C.L) in Figure~\ref{Neutrino_flux_UL} is calculated for $\nu_{\mu}$ events at $z=0.151$, for the detector geometry of IceCube Gen2 (dash–dotted curve) and GRAND200k (dotted curve).

\begin{figure}[h!]
    \centering
    \includegraphics[width = 0.8\linewidth]{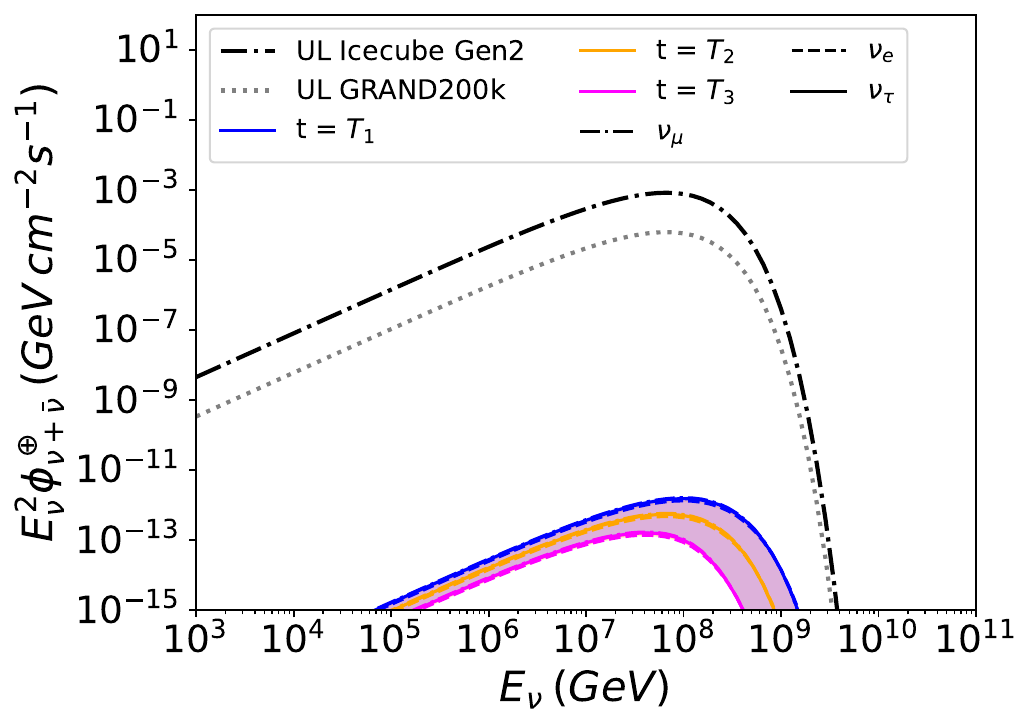}
    \caption{Neutrino Flux of GRB 221009A at three time interavals $T_{1}$, $T_{2}$ and $T_{3}$. Black Dashed-dot and grey dotted line are the upper limit sensitivity curve obtained for IceCube Gen2 and GRAND 200k for the time interval $T_{2}$.}
    \label{Neutrino_flux_GRB221009A}
\end{figure}

Now, instead of a simulated GRB event (Figures~\ref{Neutrino_flux_onoff_axis} and \ref{Neutrino_flux_UL}), we illustrate the neutrino flux of GRB~221009A at redshift $z = 0.151$ in Figure~\ref{Neutrino_flux_GRB221009A}. As discussed above, neutrino spectra are decomposed into the three neutrino flavors
$\nu_\mu$, $\nu_e$, and $\nu_{\tau}$, evaluated over successive time intervals
$T_{1}=T^*+[22,100]\,\mathrm{s}$, $T_{2}=T^*+[100,674]\,\mathrm{s}$, and $T_{3}=T^*+[674,1774]\,\mathrm{s}$ respectively. The shaded purple band corresponds to the mildly off-axis neutrino emission scenario, comprising the combined
fluxes of $\nu_{\mu}+\bar{\nu}_{\mu}$ (dashed–dotted), $\nu_{e}+\bar{\nu}_{e}$ (dashed),
and $\nu_{\tau}+\bar{\nu}_{\tau}$ (solid). Here, the neutrino spectra, similar to blue, orange, and magenta in Figure~\ref{Neutrino_flux_onoff_axis}, denote the
temporal evolution of the flux across these three time intervals. These fluxes are obtained using the same
afterglow parameters that reproduce the observed GeV–TeV emission of GRB~221009A, within a
blast wave framework expanding in a constant density ISM and evolving adiabatically.
All spectra include flavor mixing due to neutrino oscillations during propagation to Earth.
The peak flux systematically decreases with time as the external shock decelerates. We further estimated the upper limit sensitivity curve at interval $T_{2}$ for $\nu_{\mu}$ events. We can see the predicted fluxes for all three flavors remain well below the upper limit sensitivity limits of
IceCube Gen2 (dashed-dot) and GRAND200k (dashed). This result confirms the expected absence of neutrino detections from GRB~221009A with our Gaussian structured-jet afterglow model.

\section{Constraints from Afterglow Neutrino Emission}\label{section5}

In the preceding sections, we have calculated neutrino fluxes from GRB point sources and evaluated time-integrated upper limit sensitivities using the effective areas of IceCube Gen2 and GRAND200k. These results showcase that, for GRB~221009A, the expected upper limit sensitivity for $\nu_{\mu}$ events is substantially higher for GRAND200k compared to IceCube Gen2, driven by GRAND200k’s comparatively larger effective area (see Figure~\ref{Neutrino_flux_GRB221009A}). The number of detectable muon neutrino events is further determined by the differential flux at Earth, $\phi_{\nu_{\mu}}^\oplus(E_\nu)$, which depends directly on the parameters of the underlying hadronic model.

Our analysis (subsection~\ref{Neutrino_all_results}) demonstrates that both the predicted neutrino flux and its corresponding sensitivity limits are strongly influenced by key hadronic model parameters, as highlighted in Figures~\ref{Neutrino_flux_UL} and ~\ref{Neutrino_flux_GRB221009A}. Robust constraints in the feasible parameter space necessitate a systematic exploration of their correlations across a wide energy range (GeV–PeV) and temporal scales. Such comprehensive multidimensional studies are essential to identify the regions where neutrino event detections maximize and to establish realistic detection expectations for future experiments at the $90\%$ C.L searches.

\subsection{Correlations among the model parameters}

To analyze how key hadronic model parameters affect neutrino flux and event rates, we systematically examine their correlations over broad energy and timescale ranges by estimating muon neutrino events using the geometry of the GRAND200k detector, which offers a slightly larger effective area than IceCube Gen2. 

To explore the parameter space favorable to a detectable muon neutrino event, we perform simulations for those events by systematically varying the key model parameters over broad ranges motivated by afterglow modeling in the literature. We consider a simulated GRB event at a fixed cosmological redshift $z\sim0.151$. We further fix the initial bulk Lorentz factor of the jet $\Gamma_0=460$
and $k=2.5$. The other model parameters are sampled from logarithmic uniform distributions spanning optimistic intervals.
\begin{align}
E_{k,\rm iso} &\in [10^{53},10^{56}]~{\rm erg}, \nonumber \\
n_0 &\in [0.3,30]~{\rm cm^{-3}}, \nonumber \\
\epsilon_e &\in [0.01,0.3], \nonumber \\
\epsilon_B &\in [10^{-4},10^{-1}], \nonumber \\
\epsilon_p &\in [10^{-2},10^{0}], \nonumber
\end{align}

\begin{figure}[t] % single-column float
  \centering

  % Row 1
  \begin{minipage}[t]{0.49\columnwidth}
    \centering
    \includegraphics[width=\linewidth]{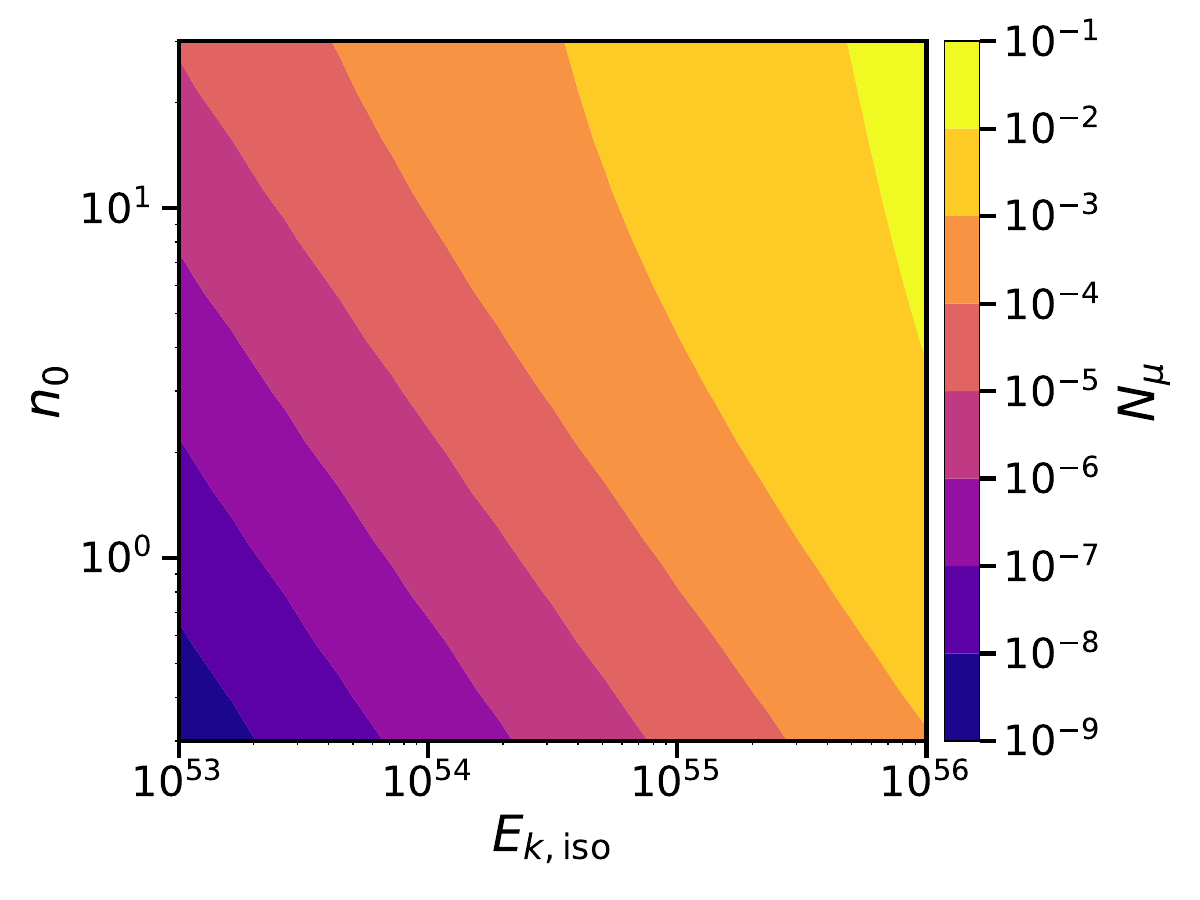}
  \end{minipage}\hfill
  \begin{minipage}[t]{0.49\columnwidth}
    \centering
    \includegraphics[width=\linewidth]{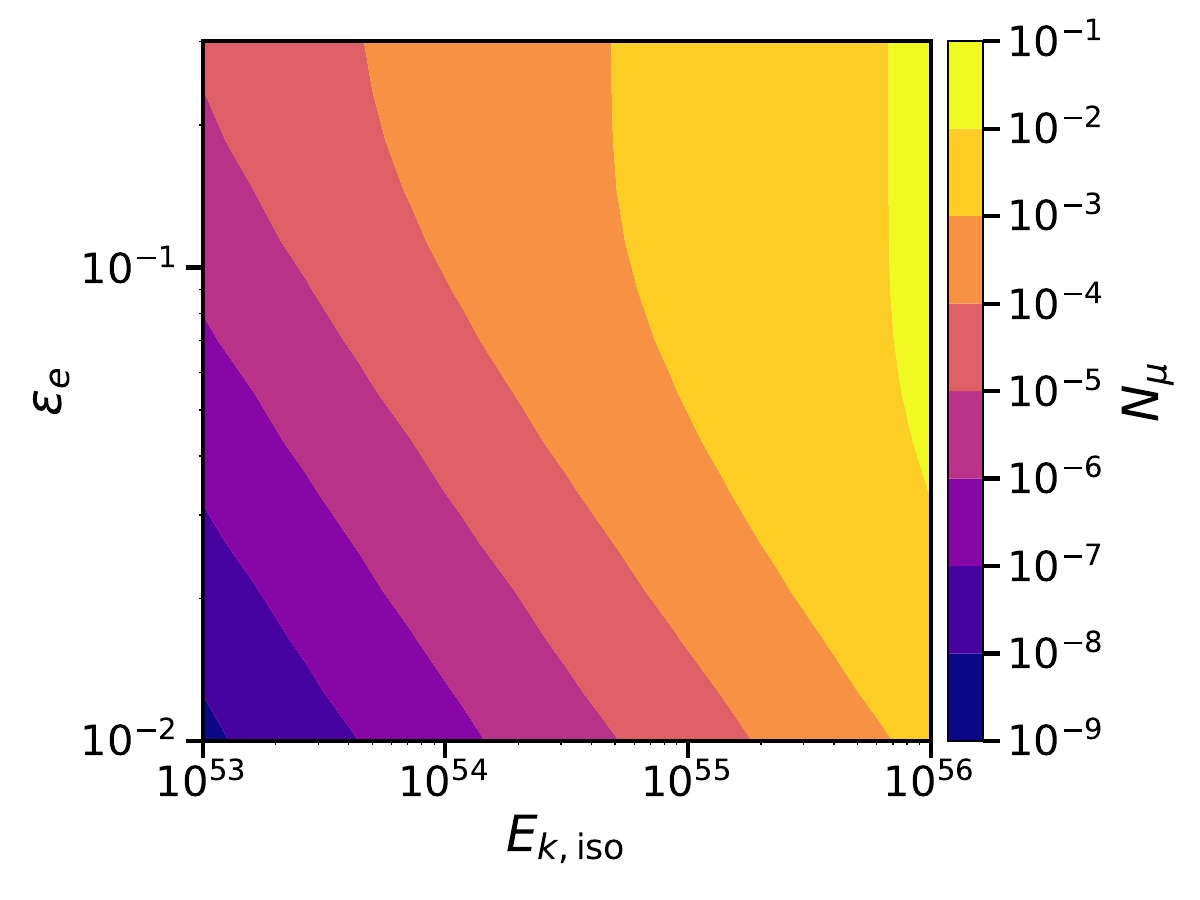}
  \end{minipage}

  \vspace{0.5em} % small vertical gap

  % Row 2
  \begin{minipage}[t]{0.49\columnwidth}
    \centering
    \includegraphics[width=\linewidth]{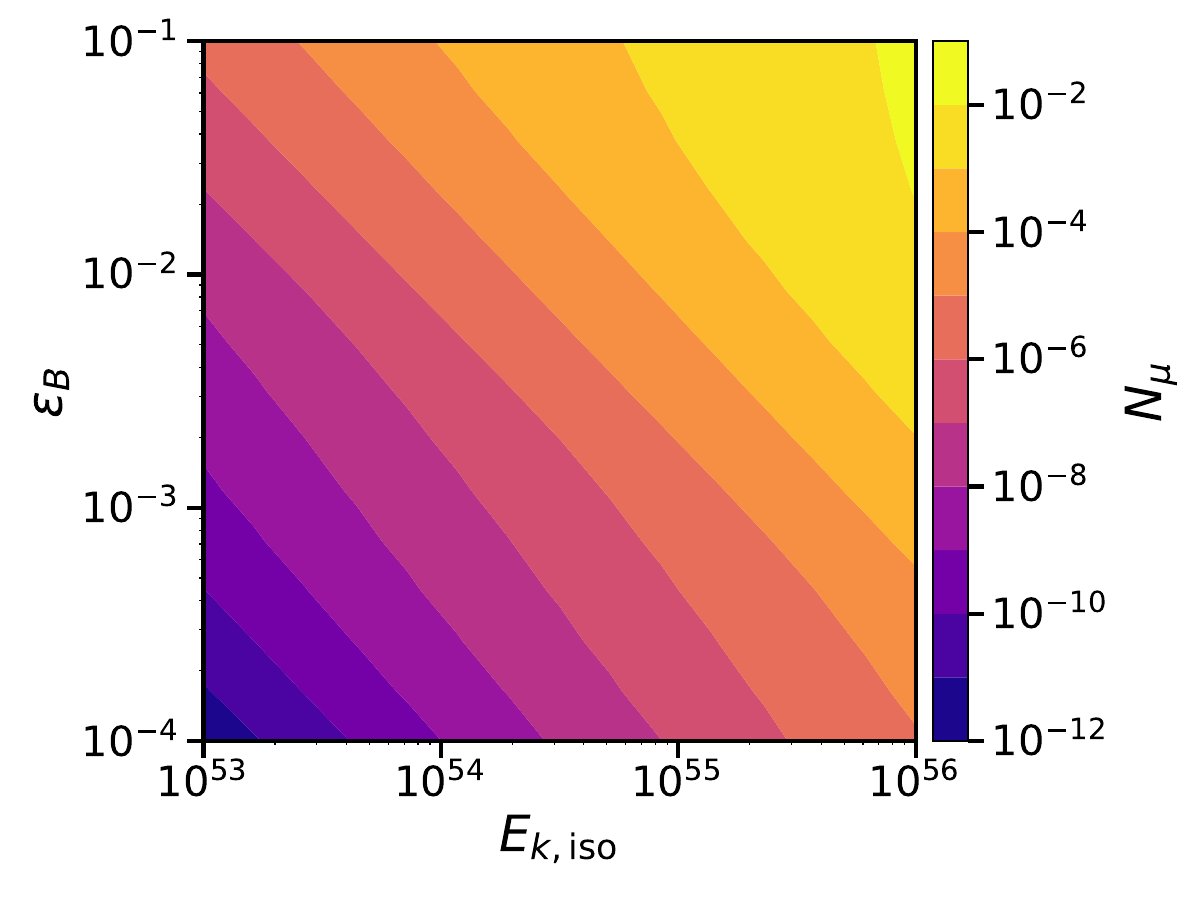}
  \end{minipage}\hfill
  \begin{minipage}[t]{0.49\columnwidth}
    \centering
    \includegraphics[width=\linewidth]{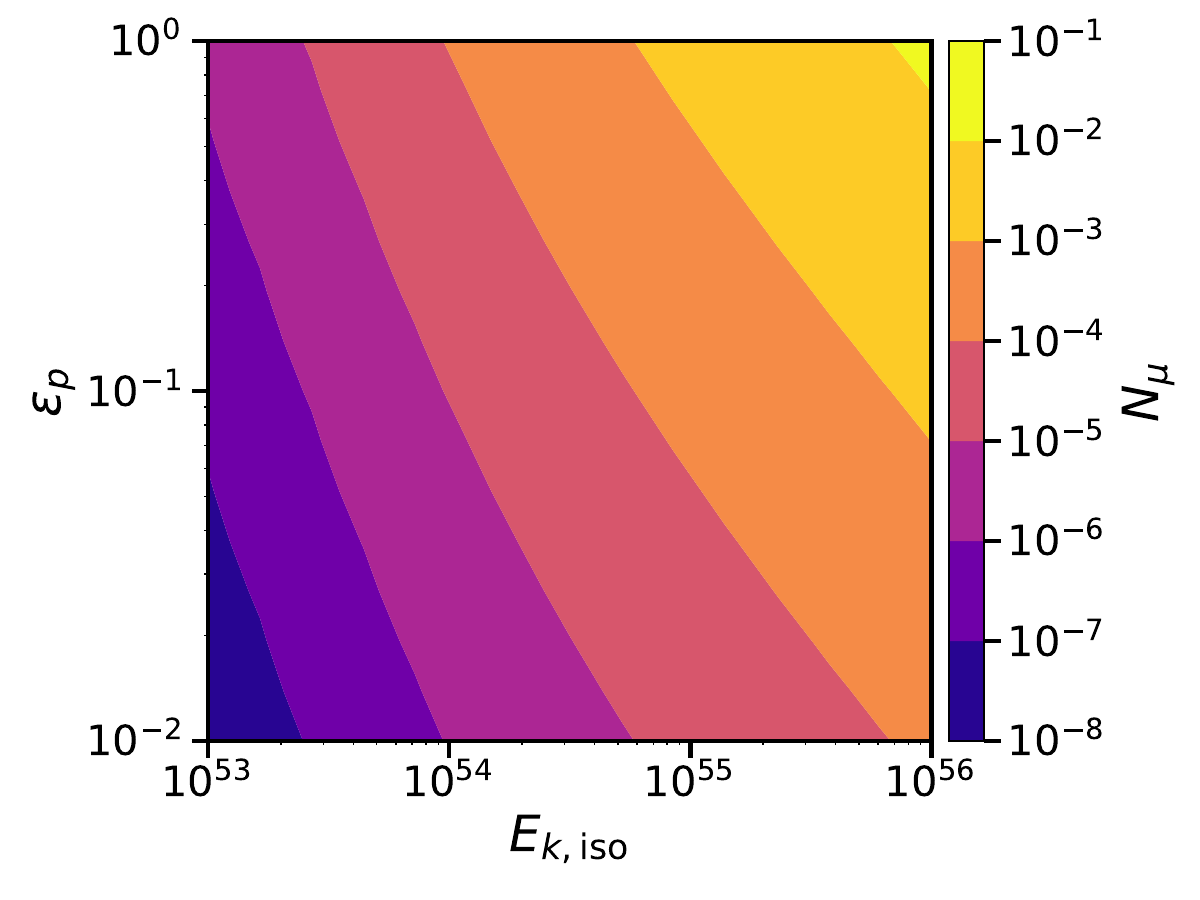}
  \end{minipage}

  \caption{Number of muon neutrino events $N_{\mu}$ for GRAND200k over $t \in \{20,2000\}\,\mathrm{s}$ as a function of $E_{k,\mathrm{iso}}$ and, panelwise, $n_0$ (top-left), $\epsilon_e$ (top-right), $\epsilon_B$ (bottom-left), and $\epsilon_p$ (bottom-right). Across all panels, $N_{\mu}$ increases monotonically with both $E_{k,\mathrm{iso}}$ and the corresponding model parameter. Within the scanned ranges, $N_{\mu}<1$, consistent with non-detection. Throughout these panels, when varying $E_{k,\mathrm{iso}}$ with a given parameter, the remaining parameters are held fixed at $\epsilon_e=0.1$, $\epsilon_B=0.1$, $\epsilon_p=1.0$, and $n_0=10$.
}
  \label{Ek_iso_with_other}
\end{figure}

\begin{figure*}[t] % two-column float; try [t] or [tbp]
  \centering

  % Three panels across full \textwidth
  \begin{minipage}[t]{0.32\textwidth}
    \centering
    \includegraphics[width=\linewidth]{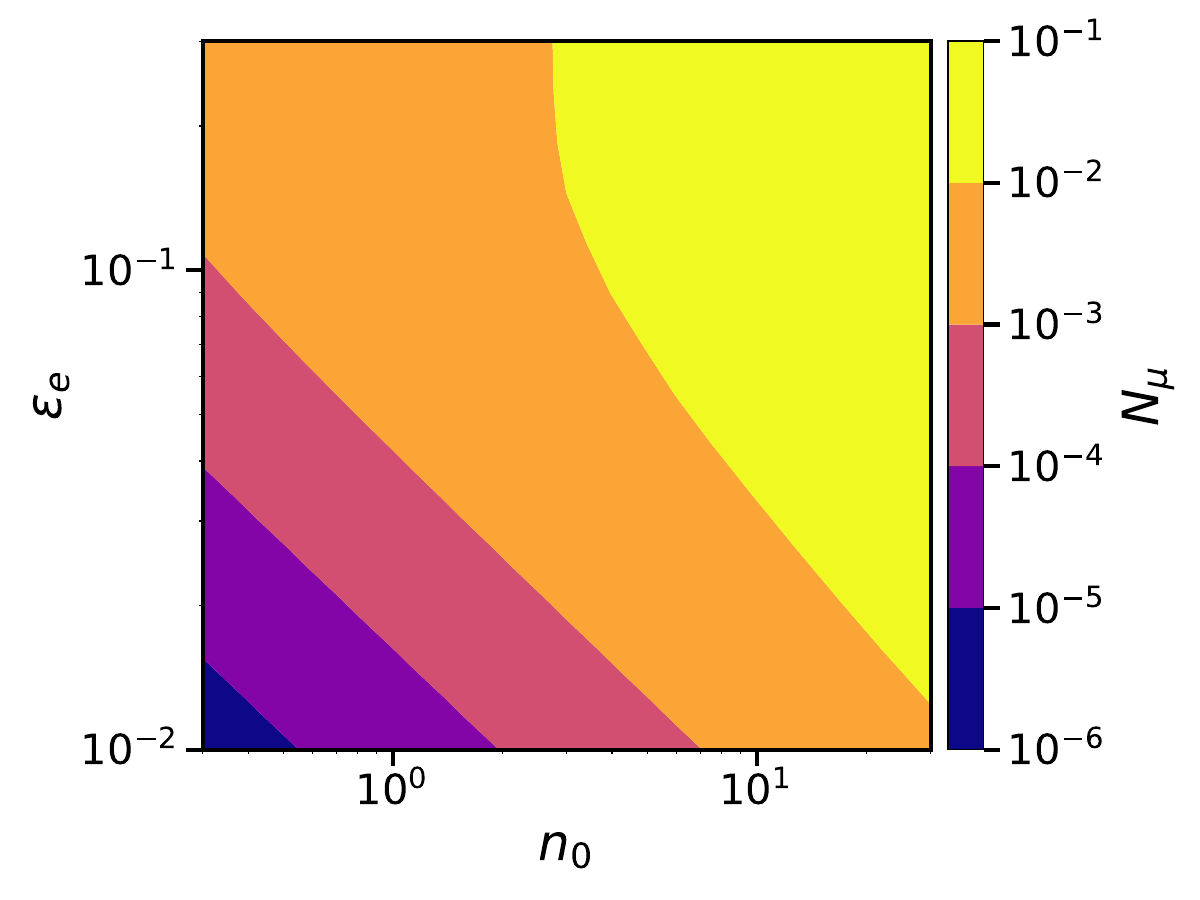}
    % \captionof{figure}{(a) Optional short note} % (avoid if you want one overall caption)
  \end{minipage}\hfill
  \begin{minipage}[t]{0.32\textwidth}
    \centering
    \includegraphics[width=\linewidth]{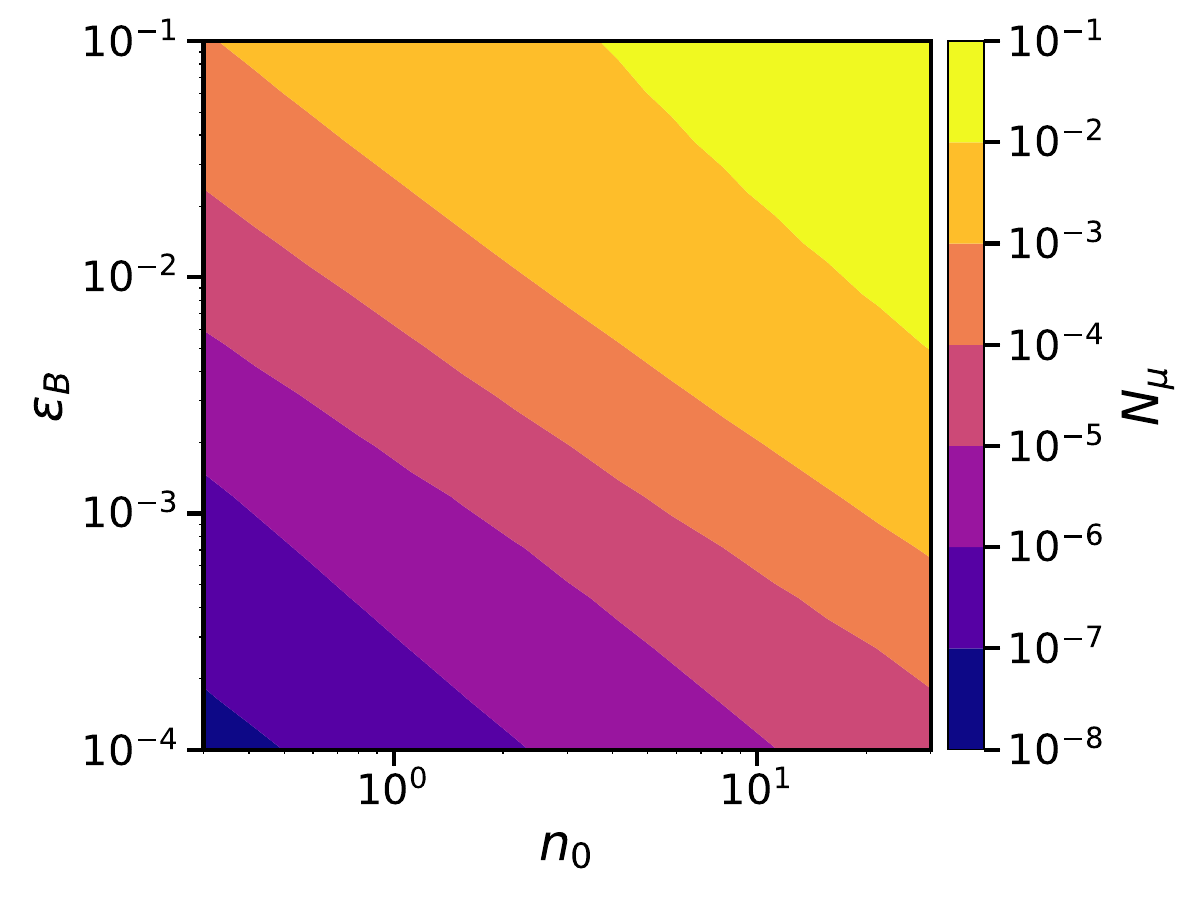}
  \end{minipage}\hfill
  \begin{minipage}[t]{0.32\textwidth}
    \centering
    \includegraphics[width=\linewidth]{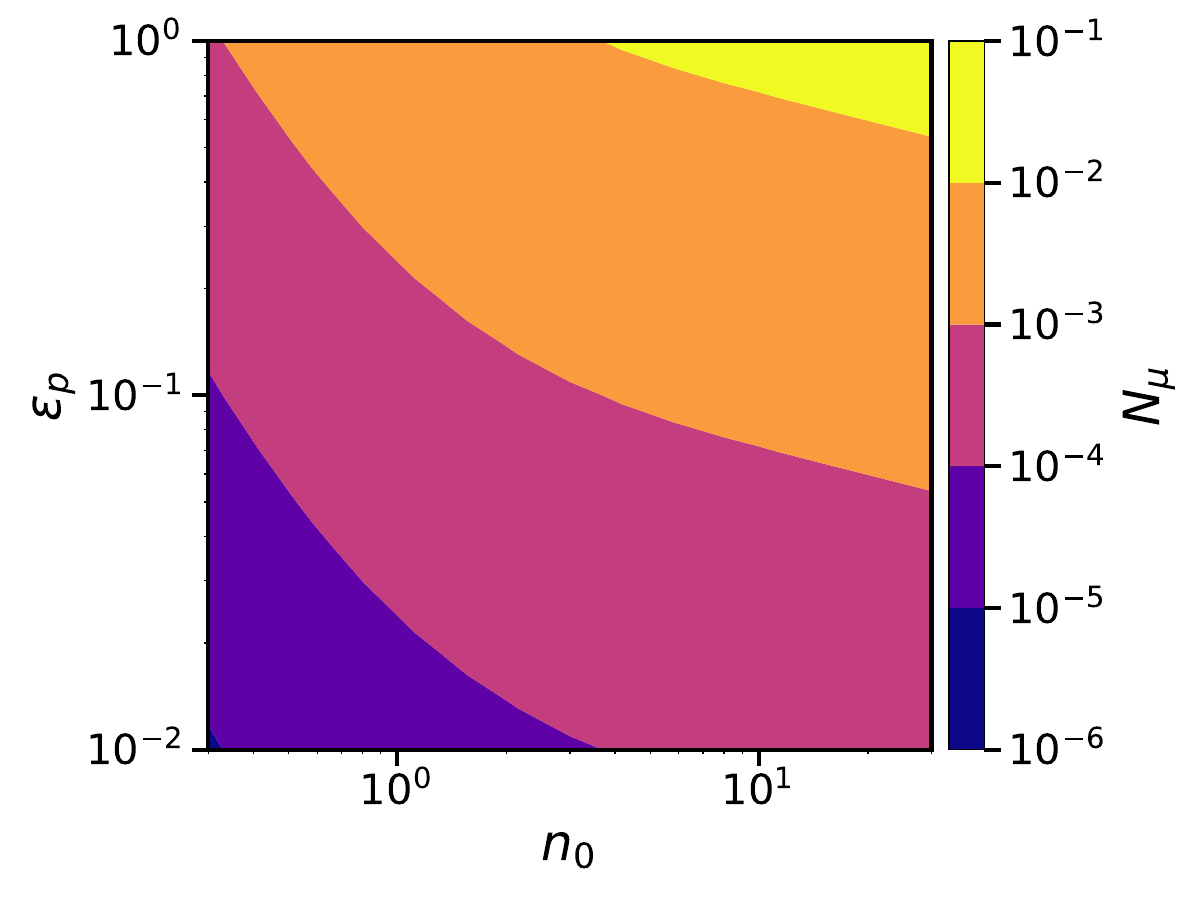}
  \end{minipage}

  \caption{Contour maps of the predicted event yield $N_{\mu}$ for GRAND200k over $t \in \{20,2000\}\,\mathrm{s}$, shown in the $n_0$–$\epsilon_e$ (left), $n_0$–$\epsilon_B$ (middle), and $n_0$–$\epsilon_p$ (right) planes. While varying one parameter with$n_{0}$, others fixed at $\epsilon_e=0.1$, $\epsilon_B=0.1$, $\epsilon_p=1.0$, and $E_{k,\mathrm{iso}}=10^{56}\,\mathrm{erg}$ accordingly. The yield grows with both $n_0$ and the associated microphysical parameter, with the strongest sensitivity along the $n_0$–$\epsilon_p$ direction.}
  \label{n0_with_others}
\end{figure*}

\begin{figure*}[t] % two-column float; try [t] or [tbp]
  \centering

  % Three panels across full \textwidth
  \begin{minipage}[t]{0.32\textwidth}
    \centering
    \includegraphics[width=\linewidth]{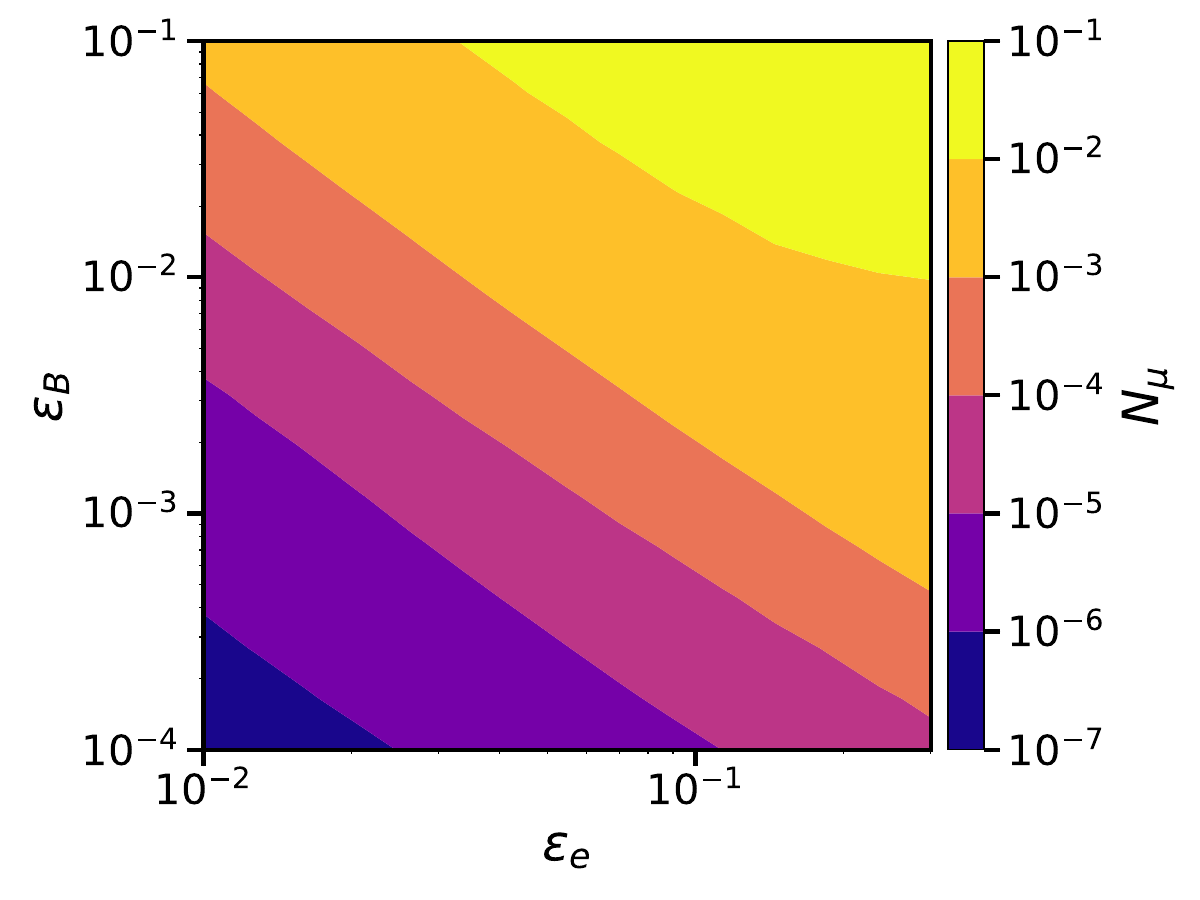}
    % \captionof{figure}{(a) Optional short note} % (avoid if you want one overall caption)
  \end{minipage}\hfill
  \begin{minipage}[t]{0.32\textwidth}
    \centering
    \includegraphics[width=\linewidth]{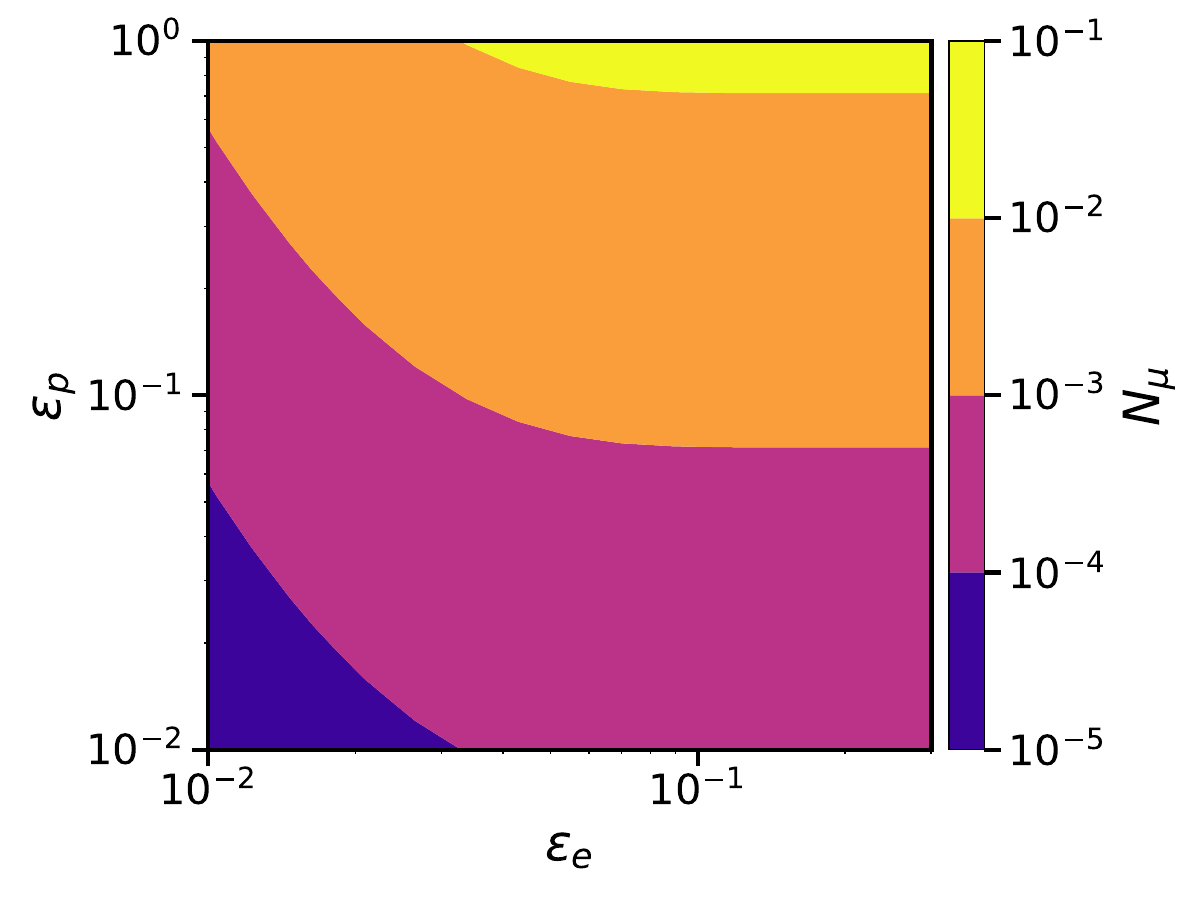}
  \end{minipage}\hfill
  \begin{minipage}[t]{0.32\textwidth}
    \centering
    \includegraphics[width=\linewidth]{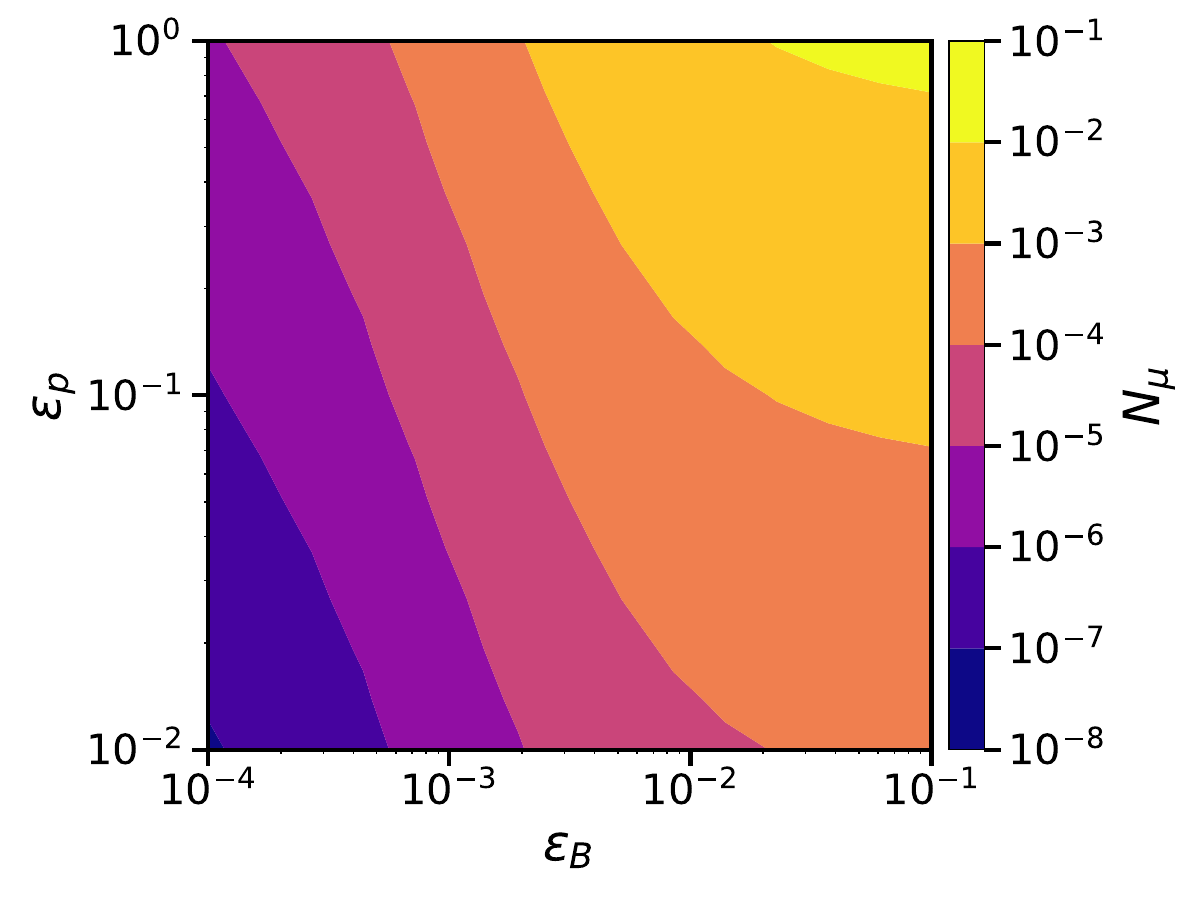}
  \end{minipage}

  \caption{Variation in the predicted GRAND200k event yield ($N_{\mu}$) as a function of microphysical parameter pairs—$(\epsilon_e,\epsilon_B)$ (left), $(\epsilon_e,\epsilon_p)$ (center), and $(\epsilon_B,\epsilon_p)$ (right), with other parameters fixed at $\epsilon_e = 0.1$, $\epsilon_B = 0.1$, $\epsilon_p = 1.0$, $n_{0}=10$ and $E_{k,\mathrm{iso}} = 10^{56}\,\mathrm{erg}$ accordingly. The contour maps show that increasing either parameter in a pair amplifies the expected neutrino event rate, with particularly strong sensitivity for joint increases in $\epsilon_e$ and $\epsilon_p$ or $\epsilon_B$ and $\epsilon_p$, underscoring the importance of electron and proton acceleration efficiencies for neutrino production.}

  \label{epse_with_others}
\end{figure*}

To investigate which parameter combinations most strongly drive detectability, we generate two-dimensional contour maps for every pair among $(E_{k,\mathrm{iso}},\, n_0,\, \epsilon_e,\, \epsilon_B,\, \epsilon_p)$, with the color scale showcasing the number of muon neutrino events (see Figures~\ref{Ek_iso_with_other},~\ref{n0_with_others},~\ref{epse_with_others}). In all panels, lighter shades correspond to larger muon neutrino events $(N_{\mu})$. The color maps show a clear trend,  increasing $E_{k,\mathrm{iso}}$, $n_0$, $\epsilon_e$, $\epsilon_B$, or $\epsilon_p$ raises the predicted $N_{\mu}$.

Physically, this reflects the fact that a more energetic blast wave ($E_{k,\rm iso} \gtrsim 10^{55}$ erg) interacting with a denser circumburst medium ($n_0 \gtrsim 1~{\rm cm^{-3}}$)  produces a larger population of shock-accelerated protons and photons, thereby boosting the efficiency of $p\gamma$ interactions. Along with a higher fraction of energy imparted to relativistic electrons ($\epsilon_e \gtrsim 1 \times 10^{-1}$), it brightens the target photon density, and leads to a larger photopion efficiency, resulting in more neutrinos. Larger magnetic fraction ($\epsilon_B \gtrsim 1 \times 10^{-3}$) further amplifies the target photon density and synchrotron cooling, which serves as the target field for neutrino production. A relatively larger proton fraction ($\epsilon_p \gtrsim 1 \times 10^{-1}$) directly boosts the non-thermal proton power available for $p\gamma$ interactions. The influence of $\Gamma_{0}$ is comparatively weak relative to the other parameters, so we keep it fixed.

For clarity, Figure~\ref{Ek_iso_with_other} presents $E_{k,\mathrm{iso}}$ scanned against $n_0$, $\epsilon_e$, $\epsilon_B$, and $\epsilon_p$; Figure~\ref{n0_with_others} shows $n_0$ versus $\epsilon_e$, $\epsilon_B$, and $\epsilon_p$; and Figure~\ref{epse_with_others} displays the microphysical variation between $(\epsilon_e,\epsilon_B)$, $(\epsilon_e,\epsilon_p)$, and $(\epsilon_B,\epsilon_p)$. In each panel, parameters not shown on the axes are held fixed at the optimistic values mentioned in the captions (e.g., when varying $E_{k,\mathrm{iso}}$ with $n_0$, we set $\epsilon_e=0.1$, $\epsilon_B=0.1$, and $\epsilon_p=1.0$; when varying $E_{k,\mathrm{iso}}$ with $\epsilon_e$, we set $n_0=10$, $\epsilon_B=0.1$, and $\epsilon_p=1.0$). Thus, these contour mappings reflect the parameter combinations that can yield comparable $N_{\mu}$ and identify the regimes most favorable for GRAND200k detectability.

% In body:

%%% Another one

\section{Summary and Conclusions}\label{section6}
%%\label{}

In this work, we modeled the very high energy afterglow of GRB~221009A with an external forward shock from a Gaussian structured jet expanding into a uniform ISM medium, and we computed the accompanying, $p\gamma$ neutrino emission in the PeV--EeV range. 

In section~\ref{section3}, we applied our Gaussian structured jet model to reproduce the GeV–TeV afterglow observed by AGILE-GRID and LHAASO to study the SSC-driven VHE afterglow features in both the spectral (SEDs Figure~\ref{Three-SEDs}) and temporal (LCs Figure~\ref{fig:LC_AGILE}) domains of GRB 221009A. We validated the model through detailed spectral fits of the GeV–TeV afterglow across three early-time intervals, accounting for KN effects, the opacity of internal $\gamma\gamma$ pair production, and EBL attenuation. We observe the resulting best-fit afterglow parameters from MCMC sampling favor a mildly off-axis jet structure. The resulting viewing angle $\theta_v=2.46^{\circ}$, and jet core angle $\theta_c=4.41^{\circ}$ drives the $E_{k,\rm iso}$ $\sim 2\times 10^{55}$ erg and here $n_{0}$ is $0.97$ $cm^{-3}$ respectively. In both SEDs and LCs, the fits require $\epsilon_e \gg \epsilon_B$, which makes SSC emission strongly dominates at VHE. To produce a consistent fit in the SEDs, we consider a time-dependent magnetic energy fraction across distinct time intervals. Slightly lower value of jet
core angle $\theta_{c}\sim 3^{\circ}$ has been proposed by  \citet{zheng2024narrow}.

The SEDs in Figure~\ref{Three-SEDs} display both synchrotron and SSC hump, with SSC dominating beyond ($\sim1$) GeV and strong suppression of EBL above ($\sim10$) TeV. Figure~\ref{fig:LC_AGILE} depicts the light curve that explains the AGILE GRID GeV data (50 MeV–3 GeV) and LHASSO TeV data (0.3–5 TeV). Our findings are consistent with independent structured-jet studies of GRB~221009A by \citet{o2023structured, gill2023grb}. However, \citet{o2023structured} primarily explains X-ray and OIR light curves and late time afterglow data with a shallow power law structured jet model, where the afterglow best-fit parameters are consistent with our results. Also \citet{gill2023grb}, explain X-ray and optical light curves of GRB 221009A with a forward shock structured jet model in a wind medium and \citet{Ren_2024} also, fit the multi-wavelength data of GRB 221009A with a structured jet geometry, taking a transition from ISM to wind-driven medium. For both cases, $E_{k, \rm iso}$ is comparable to our case.

In section~\ref{section4}, we calculate the neutrino on-axis and off-axis flux of the $p\gamma$ interaction channel. Our neutrino flux has been calculated considering the neutrino oscillation effect at Earth at time $t>t_{dec}$, and comprises three neutrino flavors $\nu_{\mu}$, $\nu_{e}$ and $\nu_{\tau}$ (see Figure~\ref{Neutrino_flux_onoff_axis}). Considering the effective areas of IceCube Gen2 and GRAND200k, we obtain neutrino event counts over a given time scale and propose a formalism to estimate the time-integrated upper limit sensitivity curve for point-like sources. A comparison of viewing geometries in Figure~\ref{Neutrino_flux_onoff_axis}, for a simulated GRB event at $z=0.151$, reveals that the on-axis geometry of the jet yields higher neutrino fluxes than the off-axis ones, due to stronger Doppler boosting and larger target photon densities. Nevertheless, even optimistic on-axis scenarios showcase in Figure~\ref{Neutrino_flux_UL} that the neutrino flux remains below the single-burst upper limit sensitivity curve. Furthermore, we estimated the neutrino flux upper limit curve of GRB 221009A (see Figure~\ref{Neutrino_flux_GRB221009A}) using the effective areas of IceCube Gen2 and GRAND200k for the time interval $T_{2}=T^*+[100,674]\,\mathrm{s}$ for muon neutrino events. The same afterglow parameters given in section~\ref{section3} are considered for neutrino flux calculations. In the Figure~\ref{Neutrino_flux_GRB221009A}, we observe that the neutrino fluxes for all three flavors remain well below the upper limit sensitivity limits of IceCube Gen2 and GRAND200k. 

Finally, in Section~\ref {section5}, we investigate how hadronic model parameters impact neutrino flux and event rates. For this purpose, we examine the model parameter correlations simulating GRB events by estimating muon neutrino events over broad energy and timescale ranges using the geometry of the GRAND200k detector. The GRAND200k detector is considered over IceCube Gen2 in this correlation study as it offers a slightly larger effective area. During the simulation, we keep $z\sim0.151$, initial bulk Lorentz factor of the jet $\Gamma_0=460$ and electron spectral index  $k=2.5$ fixed. Other model parameters are sampled from logarithmic uniform distributions spanning optimistic intervals. We find that a more energetic blast wave, interacting with a denser circumburst medium, produces a larger $N_{\mu}$ events. Further, a higher fraction of $\epsilon_e$, $\epsilon_B$ and $\epsilon_p$ enhances the neutrino production, which is shown in our Figure \ref{Ek_iso_with_other},  \ref{n0_with_others}, and \ref{epse_with_others}. However, the effect of $\Gamma_0$ is not as pronounced compared to other parameters, hence we have kept it fixed.

Our study highlights the significance of jet angular structure, particularly the on-axis and off-axis geometry, in shaping both the electromagnetic and neutrino afterglow signatures of GRBs. Furthermore, our exploration of parameter space reveals the paths forward for enhancing joint VHE and neutrino detections in future observatories, providing clear criteria for the necessary conditions for detectability. While GRBs remain viable candidates for ultra-high-energy cosmic ray sources, our results suggest that only particularly energetic, nearby, and efficiently baryon-loaded bursts are likely to produce detectable neutrino and cosmic-ray signals, thereby refining the search for the most promising multi messenger events.

\section*{Acknowledgements}
TM acknowledge the financial support received through the Prime Minister’s Research Fellowship (PMRF). TM expresses sincere gratitude to the Centre for Astro-Particle Physics (CAPP) and the Department of Physics at the University of Johannesburg, funded by a BRICS STI grant to SR, where part of this work was completed. TM is grateful to Prof.
Resmi Lekshmi (IIST, Trivandrum) for her valuable guidance and
continuous support throughout this work. TM also acknowledge the use of the Paramshakti Supercomputing Facility at IIT Kharagpur, established under the National Supercomputing Mission, Government of India, for providing the high-performance computational resources essential to this work. The authors also thank Prof. Sarira Sahu for helpful discussions and are also thankful to Sabyasachi Chakraborty (IIT Kharagpur) and Ajay Sharma (S.N.\ Bose National Centre for Basic Sciences) for further valuable discussions.

\appendix

\bibliographystyle{elsarticle-harv} 
\bibliography{GRB}

\end{document}